\documentclass[final,5p,times,twocolumn]{elsarticle}

\usepackage{amssymb}
\usepackage{amsmath}
\usepackage{graphicx}
\usepackage{subcaption}
\usepackage{booktabs}
\journal{XXX}

\begin{document}

\begin{frontmatter}

\title{Z-SSMNet: Zonal-aware Self-supervised Mesh Network for Prostate Cancer Detection and Diagnosis with Bi-parametric MRI}

%% Author name
\author[label1]{Yuan Yuan} 
\ead{yyua9990@uni.sydney.edu.au}
\author[label2]{Euijoon Ahn}
\ead{euijoon.ahn@jcu.edu.au}
\author[label1,label3]{Dagan Feng}  
\ead{dagan.feng@sydney.edu.au}
\author[label4,label5]{Mohamed Khadra}
\ead{mohamed.khadra@health.nsw.gov.au}
\author{Jinman Kim\corref{cor1}\fnref{label1,label5}}
\ead{jinman.kim@sydney.edu.au}

\affiliation[label1]{organization={School of Computer Science, Faculty of Engineering, The University of Sydney},%Department and Organization
            city={Sydney},
            postcode={2006}, 
            state={NSW},
            country={Australia}}           
\affiliation[label2]{organization={College of Science and Engineering, James Cook University},%Department and Organization
            city={Cairns},
            postcode={4870}, 
            state={QLD},
            country={Australia}}  
\affiliation[label3]{organization={Institute of Translational Medicine, Shanghai Jiao Tong University},%Department and Organization 
            city={Shanghai},
            postcode={200240}, 
            country={China}}
\affiliation[label4]{organization={Department of Urology, Nepean Hospital},%Department and Organization
            city={Sydney},
            postcode={2747}, 
            state={NSW},
            country={Australia}}
\affiliation[label5]{organization={Telehealth and Technology Centre, Nepean Blue Mountains Local Health District (NBMLHD)},%Department and Organization 
            city={Sydney},
            postcode={2750}, 
            state={NSW},
            country={Australia}}

\cortext[cor1]{Corresponding author: (jinman.kim@sydney.edu.au)}

%% Abstract
\begin{abstract}
\textbf{Background and Objective:} Bi-parametric magnetic resonance imaging (bpMRI) has become a pivotal modality in the detection and diagnosis of clinically significant prostate cancer (csPCa). Developing AI-based systems to identify csPCa using bpMRI can transform PCa management by improving efficiency and cost-effectiveness. However, current state-of-the-art methods using convolutional neural networks (CNNs) are limited in learning in-plane and three-dimensional spatial information from anisotropic images. Their performances also depend on the availability of large, diverse, and well-annotated bpMRI datasets. This study aims to develop an advanced network for anisotropic bpMRI analysis and enable the network to benefit from large-scale unlabeled data and domain-specific knowledge. \\
\textbf{Methods:} We propose a Zonal-aware Self-supervised Mesh Network (Z-SSMNet) that adaptively integrates multi-dimensional (2D/2.5D/3D) convolutions to learn dense intra-slice information and sparse inter-slice information of the anisotropic bpMRI in a balanced manner. A self-supervised learning (SSL) technique is proposed to pre-train our network using large-scale unlabeled data to learn the appearance, texture, and structure semantics of bpMRI. It aims to capture both intra-slice and inter-slice information during the pre-training stage. Furthermore, we constrained our network to focus on the zonal anatomical regions to further improve the detection and diagnosis capability of csPCa. \\
\textbf{Results:} We conducted extensive experiments on the PI-CAI (Prostate Imaging - Cancer AI) dataset comprising 10000+ multi-center and multi-scanner data. Our Z-SSMNet excelled in both lesion-level detection (AP score of 0.633) and patient-level diagnosis (AUROC score of 0.881), securing the top position in the Open Development Phase of the PI-CAI challenge and maintained strong performance, achieving an AP score of 0.690 and an AUROC score of 0.909, and securing the second-place ranking in the Closed Testing Phase. \\
\textbf{Conclusions:} Our results underscore the potential of AI-driven systems in advancing csPCa detection and diagnosis, offering a robust framework for future research and clinical application. 
\end{abstract}

%% Keywords

\end{frontmatter}



\section{Introduction}
\label{sec:introduction}
%% Labels are used to cross-reference an item using \ref command.
Prostate cancer (PCa) is the second most frequent cancer and the fifth leading cause of cancer death in men \cite{bray2024global}. Early identification of clinically significant PCa (csPCa) allows accurate diagnosis, tailored treatment, and improved patient outcomes \cite{moses2023prostate}. Bi-parametric MRI (bpMRI) has emerged as a pivotal modality in the early detection of csPCa for pre-biopsy investigations by providing 3D volumetric imaging data. It offers the advantages of shorter scan duration and reduced side effects compared to multi-parametric MRI (mpMRI) \cite{tamada2021comparison}. It has also been known to be more suitable for high-volume and population-based disease screening \cite{eklund2021mri}.

In recent years, artificial intelligence (AI)-based automatic detection and diagnosis systems have shown significant potential in enhancing the accuracy and efficacy of cancer identification \cite{sharma2023survey, iglesias2024artificial}, including the detection and diagnosis of csPCa in bpMRI \cite{sunoqrot2022artificial}. For example, Saha et al. \cite{saha2024artificial} organized the PI-CAI (Prostate Imaging: Cancer AI) challenge to develop and evaluate recent AI-based solutions for the lesion-level detection and patient-level diagnosis of csPCa in bpMRI. This has rapidly advanced the field, leveraging popular deep learning (DL) algorithms optimized for the detection of csPCa. Notably, DL approaches based on convolutional neural networks (CNNs) have achieved state-of-the-art performance. For example, Wang et al. \cite{wang2018automated} introduced an end-to-end CNN that integrated 2D T2-weighted imaging (T2W) and Apparent Diffusion Coefficient (ADC) images for the detection and diagnosis of csPCa. It jointly optimized prostate detection, T2W-ADC registration, and csPCa localization. Similarly, Cao et al. \cite{cao2019joint} developed FocalNet, a multi-class CNN that simultaneously detected csPCa and clinically insignificant PCa. In another study, Duran et al. \cite{duran2022prostattention} proposed a two-branch CNN with an attention mechanism to jointly perform multi-class segmentation of PCa lesions and prostate segmentation, where they used the prostate segmentation mask as an attention map to guide the segmentation of lesions. 

The prostate is typically partitioned into two distinct zones: the central gland (CG) (which includes both the transition zone (TZ) and the central zone (CZ)) and the peripheral zone (PZ). The PZ is notable, comprising 70\%-75\% of diagnosed cases of PCa. In contrast, 20\%-30\% of PCa cases originate in the TZ, with CZ-originating cancers being rare and often secondary to PZ invasion \cite{turkbey2019prostate}. Recently, some researchers have improved the detection of PCa lesions by learning the zone (on the image) of the PCa involvement. Vente et al. \cite{de2020deep} incorporated such zonal information into their CNN model to simultaneously detect and diagnose cancer. Similarly, Zhu et al. \cite{zhu2022fully} developed a cascaded DL model named Res-UNet for detecting csPCa using spatial constraints of the anatomical structures (CG and PZ). While these CNNs leveraged the anatomical structures, they were limited to processing only 2D individual slices and failed to capture the 3D structure of images, leading to the potential loss of spatial information and context in 3D volume.

More recently, Seetharaman et al. \cite{seetharaman2021automated} presented the Stanford Prostate Cancer Network (SPCNet). It incorporates three adjacent slices to have a 2.5D representation, aiming to better capture information between adjacent slices. This 2.5D CNN model focused on predicting cancer specifically on the middle slice and employed separate convolutional layers for each MRI sequence in bpMRI. The outputs were subsequently concatenated to indicate the presence of aggressive cancer, indolent cancer, or normal tissue. By leveraging inter-slice information, the model captured contextual relationships. However, it could not learn global 3D volumetric information.

The advent of 3D CNNs has facilitated a holistic understanding of 3D medical images. Adams et al. \cite{adams2022prostate158} conducted csPCa segmentation with a 3D U-ResNet model that captured spatial information across the entire volume. Here, a re-sampling technique was used to ensure equal spatial resolution in all image directions and maintain consistent spatial relationships. Other studies attempted to reduce the intra-slice resolution using techniques such as anisotropic pooling or stride convolution \cite{bosma2023semi}. However, these techniques introduced interpolation artifacts, leading to distortions or blurring of the original information. They were also limited in accommodating the anisotropy of the images.

All the studies mentioned earlier relied on supervised learning (SL) algorithms, and large-scale labeled data was required to train the DL models. An alternative approach is self-supervised learning (SSL), which allows one to learn and extract meaningful feature representations without manually labeling data \cite{yu2024self}. It has demonstrated remarkable progress across diverse facets of medical image analysis, including but not limited to anatomical localization \cite{frueh2022self} and disease diagnosis \cite{li2021multi}. Recently, Bolous et al. \cite{bolous2021clinically} employed U-Net and holistically-nested edge detector (HED) to extract useful semantic features from unlabeled prostate 2D MRI. They presented an SSL framework employing image restoration tasks to pre-train the U-Net and HED models. Despite improving csPCa detection, inherent loss of spatial context was acknowledged as a potential limitation in 2D images (T2W and ADC), particularly in identifying small lesions and precisely characterizing csPCa. Qian et al. \cite{qian2021procdet} presented a 3D U-Net-based model that was also pre-trained with an image restoration task. These SSL models, however, were not designed to learn the dense intra-slice information and sparse inter-slice information of anisotropic images, such as with bpMRI.

\subsection{Contributions}
\label{sec:contributions}
In this paper, we propose a Zonal-aware Self-supervised Mesh Network (Z-SSMNet) for csPCa detection and diagnosis using bpMRI. Our contributions are fourfold: \textbf{1)} Our Z-SSMNet accurately identifies csPCa lesions from indolent cancer and in a wide range of benign conditions. This is achieved by adaptively fusing multiple 2D/2.5D/3D CNNs to capture a balanced representation of dense intra-slice information and sparse inter-slice information. Our balanced representation ensures that our model can grasp the intricate details (texture and heterogeneity of lesions) and context (global anatomical structure) within the image, thereby enhancing its ability to identify lesions or abnormalities. \textbf{2)} We introduce a new SSL technique to pre-train our network using large-scale unlabeled data to learn meaningful features of csPCa lesions and capture the intra-slice and inter-slice information. \textbf{3)} We constrained our network to focus on the zonal anatomical regions further to improve the detection and diagnosis capability of csPCa. \textbf{4)} We evaluated our Z-SSMNet by participating in the PI-CAI challenge. Our Z-SSMNet was ranked first place with an Average Precision (AP) score of 0.633 and an Area Under the Receiver Operating Characteristic (AUROC) score of 0.881 in the Open Development Phase, and it achieved an AP score of 0.690 and an AUROC score of 0.909 and ranked second place in the subsequent Closed Testing Phase.

\section{Materials and Methods}
\subsection{Materials}
\label{sec:sec2.1}
We employed the PI-CAI dataset \cite{saha_2022_6517398} to detect and diagnose csPCa. Additionally, we utilized a combination of the Prostate158 dataset \cite{keno_bressem_2022_6481141}, ProstateX dataset \cite{litjens2017prostatex}, and the Medical Segmentation Decathlon (MSD) dataset \cite{antonelli2022medical} to train a zonal segmentation model to generate the zonal anatomy prior.

\subsubsection{PI-CAI Dataset}
The PI-CAI dataset includes 10000+ prostate MRI exams acquired spanning 2012 to 2021 from four different centers: Radboud University Medical Centre (RUMC), Ziekenhuis Groep Twente (ZGT), Prostaat Centrum Noord-Nederland (PCNN)) and Norwegian University of Science and Technology (NTNU). Voxel-level csPCa lesion annotations were manually delineated and/or patient-level csPCa outcomes were recorded. Each annotation was derived using all available MRI exams, diagnostic reports (radiology, pathology), and whole-mount prostatectomy specimens (if applicable). Being the largest publicly available dataset for the detection and diagnosis of csPCa to date, this dataset was divided into four parts, with the following use cases: \textbf{1) Public Training and Development Dataset.} There are 1500 cases from the centers of RUMC, ZGT, and PCNN in this cohort. Among them, 1075 cases are with benign tissue or indolent PCa, and 425 cases are with malignant lesions. The lesion-level annotations were unavailable with the benign tissue or indolent PCa. Among the 425 malignant cases, only 220 cases exhibit annotations provided by human experts. Additionally, all cases in the cohort contain voxel-level lesion delineations of csPCa generated by a report-guided CNN model proposed by Bosma et al. \cite{bosma2023semi}. \textbf{2) Private Training Dataset.} This cohort has 7607 cases from the centers of RUMC, ZGT, and PCNN. It is used exclusively by the challenge organizers. \textbf{3) Hidden Validation and Tuning Cohort.} This cohort has 100 hidden cases from the centers of RUMC, ZGT, and PCNN. It enables model selection and tuning during the Open Development Phase – Tuning. The data were registered and manually labeled by experts. \textbf{4) Hidden Testing Cohort.} This cohort has 1000 hidden cases from the centers of RUMC, ZGT, PCNN, and NTNU. It is used to test the AI algorithms in the Open Development Phase - Testing and Closed Testing Phase. All data in this cohort were also registered and manually labeled by experts. Parts of the data in this set are sourced from a disparate data center (NTNU), introducing heterogeneity in the data distribution.

The PI-CAI challenge took place in two phases: the Open Development Phase and the Closed Testing Phase. In the Open Development Phase, the models were trained on the Public Training and Development Dataset and tuned on the Hidden Validation and Tuning Cohort. The Hidden Testing Cohort was then used to determine the top 5 performing algorithms. In the Closed Testing Phase, the models were trained on the data from the Public Training and Development Dataset and Private Training Dataset and tested on the Hidden Testing Cohort.

\subsubsection{Prostate158 Dataset}
Prostate158 is a curated dataset of 3T prostate bpMRI images from 158 patients. All studies include T2W and DWI images with ADC maps. All patients were examined at a German university hospital (Charité University Hospital Berlin) between February 2016 and January 2020. Pixel-wise segmentations were provided for the CG and PZ.

\subsubsection{ProstateX Dataset}
The ProstateX dataset is a public mpMRI dataset from RUMC. Based on the original ProstateX dataset, Cuocolo et al. \cite{cuocolo2021quality} labeled the lesion masks and zonal masks of 204 cases to promote prostate lesion detection and zonal segmentation. We used these 204 labeled bpMRIs to train our zonal segmentation model.

\subsubsection{Medical Segmentation Decathlon Dataset}
MSD dataset acquired at RUMC includes 48 prostate mpMRI studies comprising T2W, DWI, and dynamic contrast-enhanced (DCE) series. Out of 48 mpMRI studies, only 32 studies are publicly available. We used a subset of two series, transverse T2W and the ADC of these 32 cases. The corresponding target regions of interest (ROIs) are the CG and PZ of the prostate.

\subsection{Overview of our Approach}
\label{sec:sec2.2}
\begin{figure*}[t]
\centerline{\includegraphics[width=\textwidth]{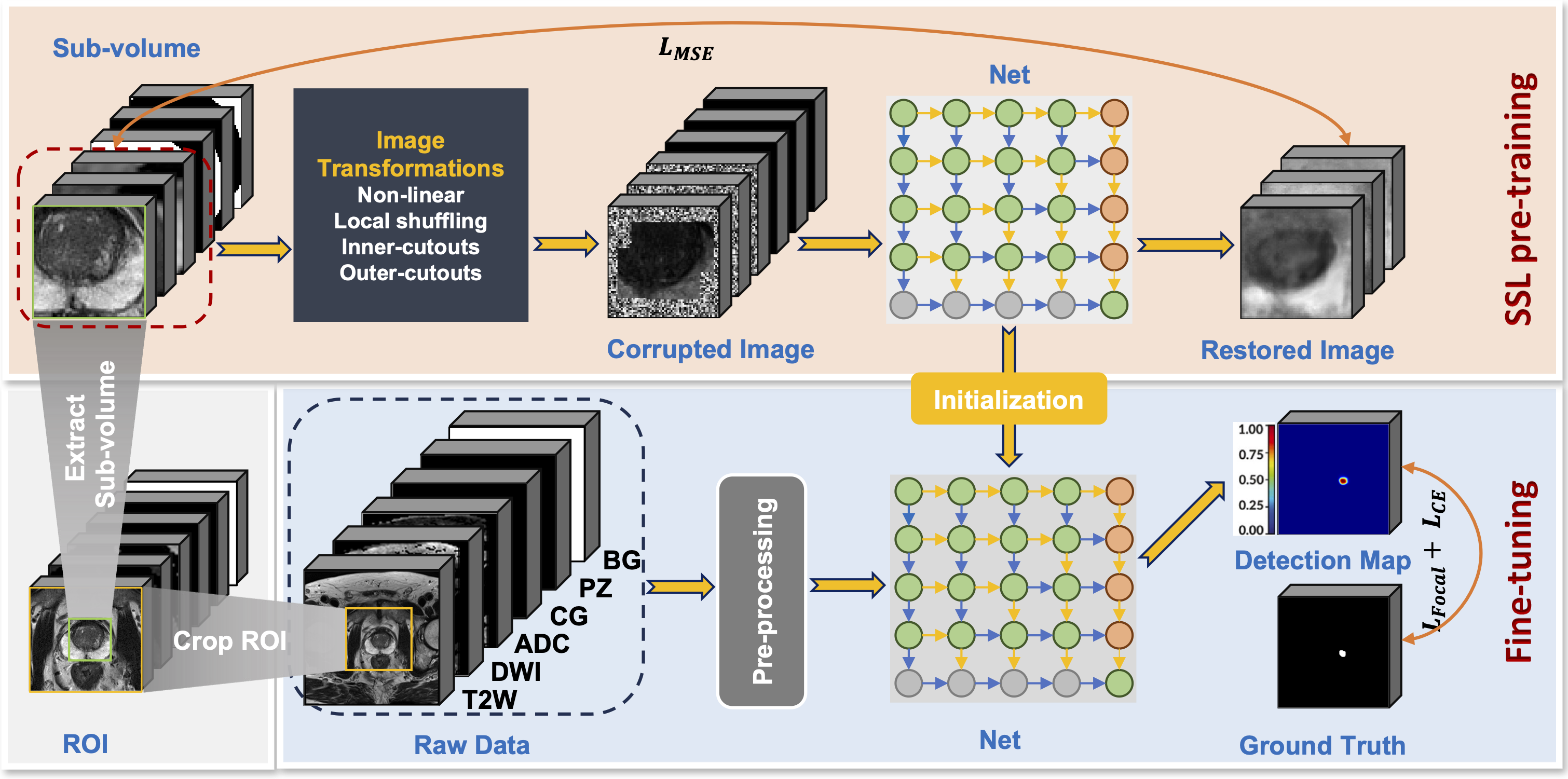}}
\caption{A schematic of our Z-SSMNet. The one-hot encoded zonal masks (CG, PZ, and BG (background)) are concatenated with the bpMRI images (T2W, DWI, and ADC) as inputs to provide anatomical priors to the model. Our SSL pre-training involves recovering the original sub-volumes of images from their corresponding corrupted images. The trained network is then fine-tuned for accurate detection and diagnosis of csPCa.}
\label{fig1}
\end{figure*}

Fig. \ref{fig1} illustrates the conceptual framework of our Z-SSMNet. Initially, the zonal masks for the CG and PZ regions were created as anatomical priors using a 3D nnU-Net trained on T2W and ADC images (More details are provided in Appendix A). The segmented zonal masks were then used in the SSL pre-training and fine-tuning stages. An image restoration pretext task was adopted to pre-train our network and was designed to acquire representations of the anisotropic information in a balanced manner. This task utilized sub-volumes of bpMRI and zonal masks as the inputs. To extract more informative regions for pre-training, we cropped the images and masks with the region of the prostate's bounding box (based on the generated zonal masks) expanding 2.5 cm outward in all directions as the ROI (Section \ref{sec:sec2.4}). Finally, We fine-tuned our network to enable accurate lesion-level detection and patient-level diagnosis of csPCa (Section \ref{sec:sec2.5}).

\subsection{Network Architecture}
\label{sec:sec2.3}
Our network comprehensively integrates traditional 2D, 2.5D, and 3D CNNs. It combines multiple CNN networks designed to accommodate images with diverse degrees of anisotropy, utilizing a multi-path approach.

\begin{figure*}[t]
\centerline{\includegraphics[width=0.8\textwidth]{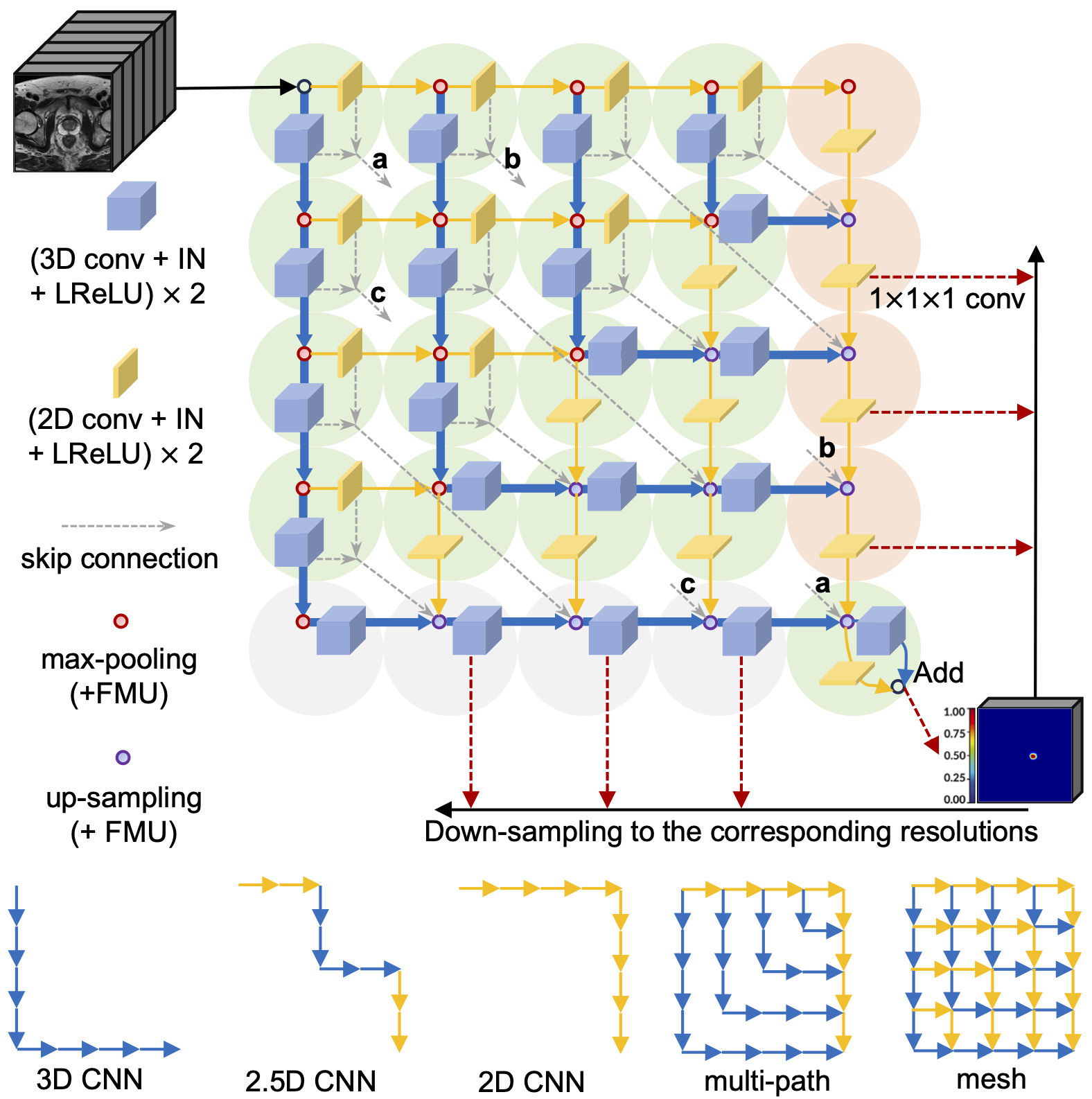}}
\caption{The network architecture of Z-SSMNet. The mesh structure concurrently combines numerous representation processes in a latent manner to dynamically create a balanced representation through adaptive learning for anisotropic information across axes. Within basic modules, both multi-dimensional and multi-level features are latently fused to harness the benefits of both 2D and 3D representations, enabling more precise modelling for target regions. Supervisory information is conveyed to six extra output branches, ensuring a thorough training of shallow layers.}
\label{fig2}
\end{figure*}

As illustrated in Fig. \ref{fig2}, our network comprises 5 $\times$ 5 (25) modules (the components inside the one circle form a module), with each module intricately connected to its adjacent counterparts. Within each module, 2D and/or 3D CNN blocks are embedded. Let us denote the entire model as $M = \left\{{m_{11}, m_{12}, ..., m_{ij}, ..., m_{55}}\right\} (i, j \in \{1,2,3,4,5\})$, where $m_{ij}$ represents the module in the $i-th$ row and $j-th$ column. Modules $m_{51}, m_{52}, m_{53}, m_{54}$ exclusively contain 3D CNN blocks, whereas modules $m_{15}, m_{25}, m_{35}, m_{45}$ only have 2D CNN blocks. The remaining modules encompass a combination of 2D and 3D CNN blocks. The 2D block comprises two $3 \times 3 \times 1$ convolution layers, whereas the 3D block consists of two $3 \times 3 \times 3$ convolution layers, both followed by instance normalization (IN) and LeakyReLU (LReLU) activation function. Our network encompasses various sub-nets with distinct structures, formed through different combinations of some modules. Any sub-net adheres to an encoding-decoding structure. In the encoding stage, max-pooling is employed to diminish the spatial resolution of the image, facilitating the integration of large-scale information. Conversely, linear interpolation in the decoding stage gradually restores the spatial resolution of the feature maps. For the encoding stage, the number of filters in each convolution block is configured as \begin{equation} K= 32 + 16 \times (D - 1) (D \in \{1,2,3,4,5\}), \end{equation} with the number of filters in the decoding stage aligning with that in the encoding stage. We follow this configuration from the MNet \cite{dong2022mnet}. Skip connections are used to counteract information loss stemming from down-sampling in the encoding stage. Initially, low-level 2D and 3D features in the encoding stage are merged through the feature merging unit (FMU), wherein corresponding elements are subtracted, and the absolute value (abs) is calculated. In the decoding stage, high-level 2D and 3D features undergo up-sampling before being merged through the FMU. The resultant merged features are concatenated. 

\subsection{Self-supervised Pre-training of the Z-SSMNet}
\label{sec:sec2.4}
Our SSL pretext task is designed to restore images that were corrupted by using image transformation techniques. Four image transformation techniques are employed: \textbf{1) Non-linear transformation.} The model is designed to restore the intensity values of an input image that has been transformed using a set of monotonic non-linear functions. To maintain the recognizability of anatomical structures, we employ Bézier Curves ~\cite{mortenson1999mathematics} with varying control points as the non-linear functions. In the process, the model learns the characteristic intensity patterns associated with normal and abnormal tissues (since certain pathologies or tissues may appear hyperintense (bright) or hypointense (dark) in comparison to surroundings), which is crucial for downstream lesion detection. The non-linear transformation also makes the model robust to imaging artifacts that may obscure or distort anisotropic information. \textbf{2) Local pixel shuffling.} We use local pixel shuffling to enhance local variations within a sub-volume while preserving global structures. Specifically, we randomly select 1000 windows for each sub-volume and sequentially shuffle the pixels within these windows. Window sizes are kept smaller than the network's receptive field to maintain the image's global content. This encourages the model to learn intricate textures and subtle details associated with lesions, leading to better discrimination between normal tissue and pathological regions. Introducing local variations through shuffling can also make the model less sensitive to imaging artifacts. \textbf{3) Inner-cutouts.} We mask the inner window regions while maintaining their surrounding context. The model acquires knowledge of local continuities in images by interpolating within each sub-volume when restoring inner-cutouts. Consistent with the recommendation by Pathak et al. \cite{pathak2016context}, the inner-cutout regions are constrained to occupy less than 1/4 of the entire sub-volume, ensuring the preservation of a challenging task. \textbf{4) Outer-cutouts.} We generate a variable quantity of windows, ranging up to 10, each with diverse sizes and aspect ratios. These windows are overlaid to create intricate shapes. When implemented on a sub-volume, the amalgamated window reveals the region inside it while masking its surroundings (outer-cutout) with a randomly determined number. The outer-cutout area is confined to occupy less than 1/4 of the entire sub-volume. Restoring outer-cutouts enables the model to learn the global geometry and spatial layout in the images through extrapolation within each sub-volume \cite{zhou2021models}. By restoring images with inner-cutouts or outer-cutouts, our mesh network adeptly captures both dense intra-slice information and sparse inter-slice information present in the anisotropic bpMRI images, achieving a balanced representation.

Our SSL process starts by extracting sub-volumes from the images, each having a size of $64 \times 64 \times 16 \times 6$. Besides the three channels in each sub-volume (T2W, DWI, and ADC images), three additional channels consist of one-hot encoded zonal masks. We randomly cropped $S$ sub-volumes from different locations of the ROI, where $S$ denotes the number of sub-volumes extracted from each case. Given $N$ samples, we extracted $S$ sub-volumes from each sample to form the new dataset ${O} = \left\{s_{1}, s_{2}, \ldots, s_{S\times N}\right\}$ ($s_{i}$ represents the $i-th$ sub-volume, $i$ $\in$ 1 to $S \times N$). Then, the application of image transformations to these sub-volumes can be expressed as,
\begin{equation}
    \widehat{O} = f(O)
\end{equation}
where $\widehat{O} = \left\{\widehat{s}_{1}, \widehat{s}_{2}, \ldots, \widehat{s}_{S \times N}\right\}$ and $f(\cdot)$ denotes an image transformation function. Subsequently, the mesh network will learn to approximate the function $g(\cdot)$, which aims to map the transformed sub-volumes $\widehat{O}$ back to their original ones $O$, that is,
\begin{equation}
    g(\widehat{O}) = O = f^{-1}(\widehat{O})
\end{equation}

Each transformation is independently applied to a sub-volume with a predefined probability, while inner-cutout and outer-cutout are considered mutually exclusive. Consequently, each sub-volume can undergo at most three of the image transformations, resulting in twelve possible transformed sub-volumes (see Appendix B).

The model converges towards the object of decreasing the mean squared error (MSE) loss, which is expressed as,
\begin{equation}
    L_{mse} = \frac{1}{S \times N} \sum_{i=1}^{i=S\times N}(g(f(s_i))-s_i)^2 
\end{equation}

\subsection{csPCa Detection and Diagnosis in bpMRI}
\label{sec:sec2.5}
After the pre-training, the network is then fine-tuned utilizing annotated data for csPCa detection and diagnosis. Given the inherent heterogeneity among data from diverse centers and vendors, we employed the same settings as the baseline nnU-Net for data pre-processing. 

We identified the issue of class imbalance during training. To mitigate this, we adopted a composite loss function that combines focal loss \cite{lin2017focal} with cross-entropy (CE) loss. The adaptive weighting mechanism of focal loss concentrates learning efforts on harder-to-classify instances, focusing the model's attention on minority classes. This is particularly advantageous when dealing with complex and diverse morphological and textural characteristics of lesions. The two losses are then averaged. The loss function is defined as,
\begin{equation}
l(Y, \widehat{Y}) = l_{focal}(Y, \widehat{Y}) + l_{ce}(Y, \widehat{Y})
\end{equation}
\begin{equation}
\begin{split}
l_{focal}(Y, \widehat{Y}) = & - \frac{1}{N}\sum_{n}(\alpha(1-\widehat{y}_{n})^{\gamma}{y}_{n}log\widehat{y}_{n} + \\ &(1-\alpha) \widehat{y}_{n}^{\gamma} (1-{y}_{n})log(1-\widehat{y}_{n}))
\end{split}
\end{equation}
\begin{equation}
l_{ce}(Y, \widehat{Y}) = - \frac{1}{N} \sum_{n} (\beta y_{n} log\widehat{y}_{n} + (1-\beta) (1-y_{n}) log(1-\widehat{y}_{n}))
\end{equation}
where $Y$ is the ground truth and $\widehat{Y}$ is the prediction. $y/\widehat{y}$ is the voxel contained in $Y/\widehat{Y}$ and $\widehat{y}$ represents the predicted probability that the voxel is malignant. $n$ denotes the index of voxels and $N$ is the number of voxels. $\alpha$ and $\gamma$ are the weighting parameters of the focal loss and $\beta$ is the weighting parameter of the cross-entropy loss.

Additionally, a thorough deep optimization strategy was implemented to fully optimize trainable parameters across varying network depths (see Fig. \ref{fig2}). We introduced six supplementary output branches using $1\times1\times1$ convolutions. Each branch's loss calculation incorporated resampled labels for corresponding detection outcomes, with the final loss derived from a weighted aggregation of individual loss components. Let $\widehat{Y}_{ij}$ denote the segmentation results of modules in row \textit{i} and column \textit{j}, and $Y_{ij}$ denotes the corresponding ground truth. The final loss can be defined as,
\begin{equation}
L = l(Y_{55}, \widehat{Y}_{55}) + \sum_{i=2}^{4}\lambda_{i}(l(Y_{i5}, \widehat{Y}_{i5}) + l(Y_{5i}, \widehat{Y}_{5i}))
\end{equation}
where $\lambda = (\frac{1}{2})^{5-i}$ is the weight of each loss item.

The Z-SSMNet yields SoftMax predictions; however, our requisition entails detection maps. We implement a dynamic lesion extraction methodology to derive lesion candidates from these SoftMax predictions, as expounded in \cite{bosma2023semi}. Subsequently, similar to the baseline methods, we utilize the maximum value of the detection map as the case-level likelihood score of harboring csPCa.

\section{Results}
\subsection{Experimental setup}
\subsubsection{Comparison with the State-of-the-Art}
We compared the performance of our Z-SSMNet with the official baselines of the PI-CAI challenge including U-Net \cite{cciccek20163d}, nnU-Net \cite{isensee2021nnu} and nnDetection \cite{baumgartner2021nndetection} and the other four top-performing AI algorithms \cite{debsdeep, kanimplementation, liprostate, karagoz2023anatomically}. U-Net \cite{cciccek20163d} has been used successfully in various medical imaging tasks due to its ability to capture both local and global contextual information. nnU-Net \cite{isensee2021nnu} expands upon the U-Net architecture by introducing a set of fixed, rule-based, and empirical parameters to enable fast, data-efficient, and holistic adaptation to new datasets for medical image segmentation tasks. Following the nnU-Net, nnDetection \cite{baumgartner2021nndetection} uses the Retina U-Net architecture \cite{jaeger2020retina}, which is designed for medical object detection. To produce the csPCa detection map, the segmentation output of the U-Net and nnU-Net were processed via the dynamic lesion extraction method described in \cite{bosma2023semi}, and the maximum value of this detection map was used as the case-level likelihood score for csPCa diagnosis. For nnDetection, the bounding box outputs were transformed into cubes with the corresponding lesion confidence, and all bounding boxes that overlapped with another bounding box of higher confidence were discarded to conform with the non-touching lesion candidates. As for the top-performing methods in the PI-CAI challenge, Debs et al. \cite{debsdeep} first segmented prostate region using nnU-Net \cite{isensee2021nnu} and then employed a second nnU-Net for voxel-level lesion segmentation on cases with csPCa and a Retina U-Net \cite{jaeger2020retina} for lesion detection. The final detection map was a linear combination of probabilities from nnU-Net, Retina U-Net, and Prostate-Specific Antigen density (PSAd). Kan et al. \cite{kanimplementation} used a 2D segmentation network named ITUNet \cite{kan2022itunet}, which combined transformers and CNNs, and a classification network named EfficientNet-b5 \cite{tan2019efficientnet}. Li et al. \cite{liprostate} developed an ensemble approach that combined predicted probability maps from the 3D U-Net \cite{cciccek20163d}, SPCNet \cite{seetharaman2021automated}, and SPCNet-Decision models. SPCNet-Decision incorporated SPCNet for voxel-level cancer detection and utilized a decision head classifier for slice-level classification. Karagöz et al.\cite{karagoz2023anatomically} incorporated the prior components of probabilistic prostate gland segmentation (CG and PZ) into a nnU-Net-based model. They also utilized clinical indices, including PSA levels and ADC intensity distributions in lesion regions, to reduce false positives.

We conducted a comprehensive set of experiments using different subsets of the PI-CAI dataset (described in Section \ref{sec:sec2.1}): \textbf{Set 1} – Fully manual labeled PI-CAI Public Training and Development Dataset including 1295 cases (supervised setting). \textbf{Set 2} – Partly manually labeled PI-CAI Public Training and Development Dataset including 1500 cases (1295 cases that have been annotated manually + 205 cases that have undergone labeling through AI techniques - semi-supervised setting). \textbf{Set 3} – PI-CAI Hidden Validation and Tuning Cohort. \textbf{Set 4} – PI-CAI Hidden Testing Cohort. \textbf{Set 5} – PI-CAI Public Training and Development Dataset combined with the Private Training Dataset.

We assessed baseline algorithms trained on Sets 1 and 2 and tested them on Sets 3 and 4, respectively. Additionally, we evaluated our Z-SSMNet and other top-performing algorithms trained on Set 2 and tested on Sets 3 and 4. The challenge organizers further retrained these models on Set 5 and tested them on Set 4, with baseline algorithms trained on the datasets using supervised and semi-supervised settings, respectively.

\subsubsection{Ablation Study}
To measure the effectiveness of our balanced representation (combined 2D, 2.5D, and 3D), the SSL pre-training, and the zonal anatomic prior, we conducted an ablation study where we progressively integrated these modules into the base network: \textbf{1) Baseline}: the nnU-Net as defined by the PI-CAI challenge organizers. \textbf{2) Baseline + mesh network (nnMNet)}: our mesh network architecture replaces the conventional U-Net. \textbf{3) Baseline + mesh network + SSL (SSMNet)}: addition of SSL pre-training of nnMNet followed by fine-tuning. \textbf{4) Baseline + mesh network + zonal masks (Z-nnMNet)}: addition of the zonal priors as ancillary information.
We utilized 5-fold cross-validation on Set 2 for model evaluation. Only cases with manual expert labels were included in calculating the evaluation metrics.

\subsubsection{Evaluation Metrics}
To evaluate the models' performance in csPCa detection and diagnosis, we used the "hit criterion" from the PI-CAI challenge, aligning with \cite{saha2021end}. Lesion-level detection was measured using the AP score, while patient-level diagnostic performance was assessed with the AUROC metric. The overall ranking of each AI algorithm in the PI-CAI challenge was based on the average of these metrics.

\subsubsection{Implementation Details}
Our Z-SSMNet was implemented on an NVIDIA GeForce RTX 3090 GPU using PyTorch. We re-sampled the images to the spacing of 3 mm $\times$ 0.5 mm $\times$ 0.5 mm by following the official pre-processing techniques\footnote{https://github.com/DIAGNijmegen/picai\_prep} suggested by the PI-CAI challenge organizers. For the SSL pre-training stage, we adopted a stochastic gradient descent (SGD) optimizer with a momentum of 0.9. A batch size of 24 was employed for training, and the initial learning rate of 0.1 was gradually decreased in line with the step learning rate policy (decays the learning rate of each parameter group by gamma (multiplicative factor of learning rate decay) every step\_size epochs). A convergence criterion was established, whereby training would halt if no decrement in the loss function was observed within 20 epochs. Transitioning to csPCa detection and diagnosis, we adhered to the five-fold cross-validation partitioning stipulated by the PI-CAI challenge organizers. During training, the learning rate initialization at 0.01 gradually diminished in accordance with the “poly” learning rate policy ($1-epoch/epoch_{max}$) \cite{chen2017deeplab}. The training utilized a patch size of $320\times320\times16$ and a batch size of 2. In inference mode, an equally weighted ensemble of the five models generated the voxel-wise output lesion map. Our code is publicly available at https://github.com/yuanyuan29/Z-SSMNet.

\subsection{Model Performance}
\subsubsection{Comparison with the State-of-the-Art}
The evaluation results using Set 3 of models trained on Set 1 or Set 2 are shown in Table \ref{table1}. Our model was ranked second and achieved an AP score of 70.9\% and an AUROC score of 89\%. Hevi AI's model achieved the highest AP score, 2.3\% higher than ours, while the AUROC score is 0.2\% lower than ours. Swangeese's model achieved the highest AUROC score, 2.8\% higher than ours, while the AP score is 6\% lower than ours. All the methods developed using Set 2 (semi-supervised setting) performed better than the baseline models trained using only Set 1 (supervised setting), both for lesion-level detection and patient-level diagnosis.

\begin{table}[t]\scriptsize
\centering
\begin{tabular}{cccc}
\toprule
\textbf{Algorithms}         & \textbf{Score} & \textbf{AUROC} & \textbf{AP} \\
\midrule
nnDetection (supervised)        & 0.502          & 0.735          & 0.269  \\
U-Net (supervised)              & 0.576          & 0.689          & 0.463  \\
nnU-Net (supervised)            & 0.597          & 0.737          & 0.457  \\
nnU-Net (semi-supervised)     & 0.714          & 0.818          & 0.610    \\
U-Net (semi-supervised)       & 0.731          & 0.829          & 0.633    \\
nnDetection (semi-supervised) & 0.734          & 0.885          & 0.582    \\
DataScientX \cite{debsdeep}   & 0.761          & 0.866          & 0.655    \\
PIMed-Stanford \cite{liprostate} & 0.783          & 0.872          & 0.693 \\
Swangeese \cite{kanimplementation} & 0.784          & 0.918        & 0.649 \\
\textbf{Z-SSMNet (Ours)}        & 0.800          & 0.890          & 0.709  \\
Hevi AI \cite{karagoz2023anatomically}           & 0.810          & 0.888          & 0.732       \\
\bottomrule
\end{tabular}
\caption{Evaluation results on Set 3. The models were trained on Set 1 or Set 2.}
\label{table1}
\end{table}

The test results using Set 4 of models trained on Set 1 or Set 2 are shown in Table \ref{table2}. Our model achieved the highest score compared to other methods, with the highest AP score of 63.3\% and AUROC score of 88.1\%. Our AUROC score is 0.8\% lower than DataScientX's and Hevi AI's. All the methods developed using Set 2 showed better performances than the baseline models developed using Set 1, both for lesion-level detection and patient-level diagnosis. We visualized the lesion detection results of the baseline methods, the method developed by Kan et al. \cite{kanimplementation}, and our Z-SSMNet trained on Set 2 in Fig. \ref{fig3}. The lesion-level PR and FROC curves and the patient-level ROC curve are presented in Appendix C. The visual results demonstrate the superior detection performance of our model.

\begin{table}[t]\scriptsize
\centering
\begin{tabular}{cccc}
\toprule
\textbf{Algorithms}      & \textbf{Score} & \textbf{AUROC} & \textbf{AP} \\
\midrule
nnDetection (supervised)    & 0.586          & 0.785          & 0.386       \\
nnU-Net (supervised)        & 0.626          & 0.803          & 0.450       \\
U-Net (supervised)          & 0.635          & 0.814          & 0.456       \\
nnDetection (semi-supervised) & 0.710          & 0.874          & 0.546     \\
U-Net (semi-supervised)       & 0.712          & 0.848          & 0.576     \\
nnU-Net (semi-supervised)     & 0.721          & 0.865          & 0.576     \\
Swangeese \cite{kanimplementation} & 0.740          & 0.886         & 0.593 \\
PIMed-Stanford \cite{liprostate} & 0.742          & 0.871          & 0.612  \\
Hevi AI \cite{karagoz2023anatomically} & 0.752    & 0.889          & 0.614  \\
DataScientX \cite{debsdeep}   & 0.752          & 0.889          & 0.615     \\
\textbf{Z-SSMNet (Ours)}      & 0.757          & 0.881          & 0.633     \\
\bottomrule
\end{tabular}
\caption{Test results on Set 4. The models were trained on Set 1 or Set 2.}
\label{table2}
\end{table}

\begin{figure*}[t]
\centerline{\includegraphics[width=\textwidth]{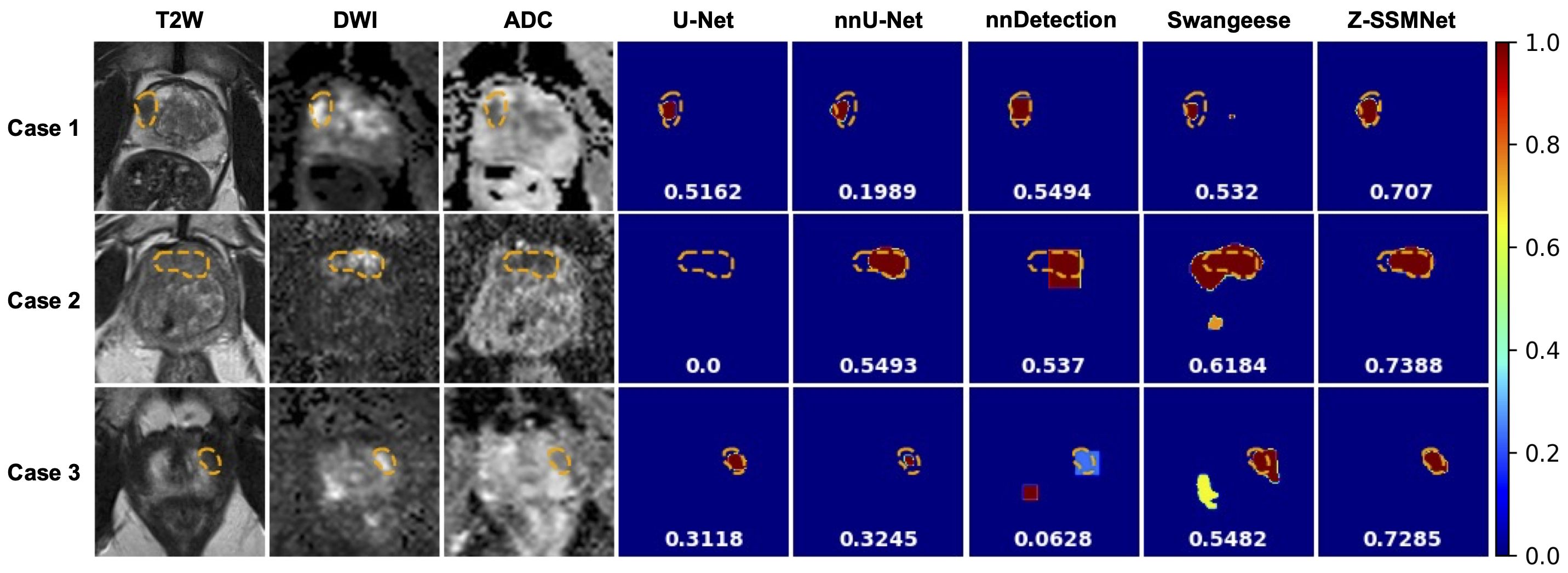}}
\caption{Visualization of csPCa detection maps from our proposed Z-SSMNet and comparing methods. The detection maps and corresponding T2W, DWI, and ADC scans (columns) are shown for three cases (rows). All the methods were trained on Set 2 (semi-supervised setting). The numbers corresponding to the bottom of the images represent the maximum probability value of the detection map.}
\label{fig3}
\end{figure*}

The separate test results provided by the PI-CAI challenge organizers in the Closed Testing Phase are shown in Table \ref{table3}, where the models were trained on Set 5 and tested on Set 4. Our Z-SSMNet was ranked second with an AP score of 0.3\% and an AUROC score of 2.5\% lower than the model of DataScientX. Here, nnDetection trained on the datasets with only manually labeled data performed better than the other settings of baseline methods.

\begin{table}[t]\scriptsize
\centering
\begin{tabular}{cccc}
\toprule
\textbf{Algorithms}       & \textbf{Score} & \textbf{AUROC} & \textbf{AP} \\
\midrule
U-Net (supervised)        & 0.553          & 0.755          & 0.351       \\
U-Net (semi-supervised)   & 0.594          & 0.773          & 0.416       \\
nnU-Net (supervised)      & 0.718          & 0.878          & 0.558       \\
nnDetection (semi-supervised) & 0.718          & 0.867          & 0.569   \\
nnU-Net (semi-supervised)     & 0.734          & 0.881          & 0.586   \\
nnDetection (supervised)      & 0.749          & 0.891          & 0.606   \\
Hevi AI \cite{karagoz2023anatomically} & 0.754   & 0.892        & 0.616   \\
PIMed-Stanford \cite{liprostate}  & 0.779        & 0.900        & 0.659   \\
Swangeese \cite{kanimplementation} & 0.795       & 0.904        & 0.686   \\
\textbf{Z-SSMNet (Ours)}        & 0.800          & 0.909        & 0.690   \\
DataScientX \cite{debsdeep}     & 0.814          & 0.934        & 0.693   \\
\bottomrule
\end{tabular}
\caption{Test results on Set 4. The models were trained on Set 5.}
\label{table3}
\end{table}

\subsubsection{Ablation Study}
Table \ref{table4} shows the improvements in detection and diagnosis accuracy due to the use of the mesh network, the SSL pre-training, and the zonal prior information. The results show that the mesh network provided a substantial improvement of 3.56\% in the AP score and 1.91\% in the AUROC score. The addition of SSL pre-training contributed to improvements of 0.94\% in AP score and 0.1\% in AUROC score. Further progress was achieved with zonal prior information, resulting in an increase of 1.45\% in AP score. Combining all the modules elevated the AP by 5.14\% and AUROC by 2.38\% compared to the baseline. The models performed better on center-cropped images than on uncropped ones.

% \scriptsize
\begin{table*}[t]
\centering
\setlength{\tabcolsep}{2mm}
\begin{tabular}{lcccccc}
\toprule
                                   & \multicolumn{3}{c}{\textbf{Test on original images}}                    
                                   & \multicolumn{3}{c}{\textbf{Test on central cropped images}}                             \\
                                   \hline\hline
                                   & \multicolumn{1}{c}{Score} & \multicolumn{1}{c}{AUROC} & \multicolumn{1}{c}{AP}  &
                                   \multicolumn{1}{c}{Score} & \multicolumn{1}{c}{AUROC} & \multicolumn{1}{c}{AP} \\
                                   \midrule

nnUNet (baseline)    & 0.6295 & 0.8352 & 0.4237 & 0.6334 & 0.8348 & 0.4320  \\
nnMNet               & 0.6569 & 0.8543 & 0.4593 & 0.6591 & 0.8518 & 0.4663  \\
nnMNet + SSL         & 0.6620 & 0.8553 & 0.4687 & 0.6649 & 0.8562 & 0.4736  \\
nnMNet + zonal mask  & 0.6641 & 0.8542 & 0.4738 & 0.6674 & 0.8559 & 0.4789  \\
nnMNet + zonal mask + SSL & \textbf{0.6671}   & \textbf{0.8590} & \textbf{0.4751} & \textbf{0.6699}  & \textbf{0.8579}   & \textbf{0.4818} \\

\bottomrule
\end{tabular}
\caption{Ablation study on Set 2 using cross-validation.}
\label{table4}
\end{table*}

\section{Discussion}
Our main findings are that: 1) Our Z-SSMNet outperformed baseline methods including U-Net \cite{cciccek20163d}, nnU-Net \cite{isensee2021nnu} and nnDetection \cite{baumgartner2021nndetection} that were trained using datasets in both supervised and semi-supervised settings for lesion-level detection and patient-level diagnosis of csPCa. Our method also had competitive performances with the state-of-the-art algorithms using datasets in semi-supervised settings; 2) Our Z-SSMNet with the mesh network was able to learn the fine-grained local representations and contextual understanding of 3D anisotropic medical images; 3) Our Z-SSMNet with the SSL pre-training captured the underlying 2D/3D structure and semantics presented in the label-free anisotropic data and, 4) Our Z-SSMNet was able to leverage the unique characteristics of different anatomical zones to improve the detection and diagnosis of csPCa.

Our model showed the best performance during the Open Development Phase - Testing of the PI-CAI challenge (Table \ref{table2}), even though its performance was second to Hevi AI's model during the Open Development Phase - Tuning (Table \ref{table1}). This demonstrates that our Z-SSMNet could generalize better across the datasets compared to other methods. The observed performance decline in Hevi AI's model, when evaluated on a dataset inclusive of cross-center data, may stem from potential overfitting to the training center-specific data. Hevi AI employed clinical variables, specifically PSAd and ADC intensity distributions, to mitigate false positive and false negative predictions. Notably, the empirically set thresholds for PSAd and ADC intensity were derived from the PI-CAI Public Training and Development Dataset, potentially contributing to the overfitting challenge. The outcomes of the Closed Testing Phase (Table \ref{table3}) further underscore their model's limited adaptability to variations in data distribution. The AUROC score achieved by our model during the Open Development Phase - Testing (Table \ref{table2}) is 0.8\% lower than that of Hevi AI and DataScientX. We attribute this to the use of clinical variables (PSAd and/or ADC intensity distributions). Moreover, the model of DataScientX achieved a 2.5\% higher AUROC score compared to ours during the Closed Testing Phase (Table \ref{table3}), which, in addition to incorporating clinical variables, also benefited from the ensemble of segmentation and detection networks. The model developed by Swangeese also integrates a segmentation network (ITUNet) and a classification network (EfficientNet-b5), allowing it to achieve competitive diagnostic results during three tests (Table \ref{table1}, \ref{table2}, \ref{table3}). However, PIMed-Stanford's model with 2.5D and 3D network integration only achieved sub-optimal results, indicating the need to optimize the network itself. The results from the challenge organizers \cite{saha2024artificial} also show that the ensemble of numbers of excellent models can further improve the results. Our model achieved an AP score of only 0.3\% lower than the best method (DataScientX) in the Closed Testing Phase (Table \ref{table3}), which further proves the excellent detection performance of our Z-SSMNet.

We evaluated the contributions of our mesh network architecture and found that it improved both lesion-level detection and patient-level diagnosis when using it to replace the U-Net architecture (Table \ref{table4}). This supports our hypothesis that combining multi-dimensional information and considering the anisotropy is essential for csPCa detection and diagnosis. Our model also exhibited improved detection accuracy and false positive reduction (Fig. \ref{fig3}, Appendix C), supporting that our mesh network had better capability in understanding fine-grained local representations and contextual relationships of 3D anisotropic medical images where our mesh network was able to capture the relations of lesions to surrounding anatomical structures and identify patterns and anomalies. 

Our SSL pre-training also contributed to improving the detection and diagnosis of csPCa compared to the models with randomly initialized weights (Table \ref{table4}). This result demonstrates that pre-training is an important tool for capturing rich and meaningful features from large, multi-center prostate bpMRI datasets without labels, and achieving effective transfer between data centers. Furthermore, our SSL pretext task was integrated into the mesh network, thus allowing the network to better learn the underlying anatomical structure and heterogeneous semantics of anisotropic bpMRI in the pre-training stage. Our network also benefited from different aspects (e.g., intensity distributions, local textures, and global structures) simultaneously by unifying the four image transformations into one pretext task.

Our results (Table \ref{table4}) also indicate that encoding the anatomical prior into our Z-SSMNet can leverage domain-specific clinical knowledge to guide the csPCa detection and diagnosis. The importance of using zonal information is consistent with prior works, e.g., Saha et al. \cite{saha2021end} and also in how radiologists rely on zonal information when making diagnostic decisions \cite{turkbey2019prostate}. Hevi AI also integrated the zonal prior to their model and improved their results. However, their AP score is lower than ours when testing (Table \ref{table2} and \ref{table3}), demonstrating the superiority of our mesh network and SSL pre-training.

\subsection{Limitations and Future Work}
We identified a few limitations of our work and potential future research opportunities. 1) We only focused on MRI without incorporating clinical data of the patients, such as PSA level, PSAd, and age of patients. Adding such clinical data to our Z-SSMNet as in \cite{debsdeep} is likely to provide further improvements. We will explore this in our future work. 2) In this study, we assumed that the errors from image registration of different modalities were adequate; however, there may be inaccurate registration results in the dataset, which may have led to sub-optimal results. In future work, we will explore registration techniques that are optimized for prostate mpMRI, for example, Giannini et al. \cite{giannini2015novel}, to further improve our Z-SSMNet for csPCa detection and diagnosis.

\section{Conclusion}
Our research presents the Z-SSMNet model, a novel approach for detecting and diagnosing csPCa using bpMRI. By integrating multi-dimensional features through deep latent fusion of 2D/2.5D/3D convolutions, our model achieves a balanced synthesis of information crucial for accurate diagnosis. We introduced self-supervised pre-training to enhance the model's understanding of fine-grained image details and contextual nuances without reliance on manual labels, alongside a novel fusion strategy that incorporates zonal anatomical priors to refine domain-specific learning. Our findings significantly surpass those of existing methods, evidenced by our top-ranking performance with a score of 0.757 in the PI-CAI challenge's Open Development Phase. Subsequently, in the Closed Testing Phase, our model continued to demonstrate strong performance, achieving a competitive score of 0.800. By pushing the boundaries of current technology, our approach paves the way for more accurate and reliable diagnostic tools, ultimately benefiting patients and clinicians alike in the fight against prostate cancer.

\section*{Acknowledgments including declarations}
This work was supported by the Australian Research Council (ARC) Grant [grant number DP200103748]. The authors also gratefully acknowledge the PI-CAI challenge organizers and sponsors for organizing the challenge. 

The authors declare that they have no known competing financial interests or personal relationships that could have appeared to influence the work reported in this paper. The ethical approval is not applicable.

\appendix
\section{Zonal Mask Generation}
\label{sec1}
\begin{figure*}[]
\centerline{\includegraphics[width=\textwidth]{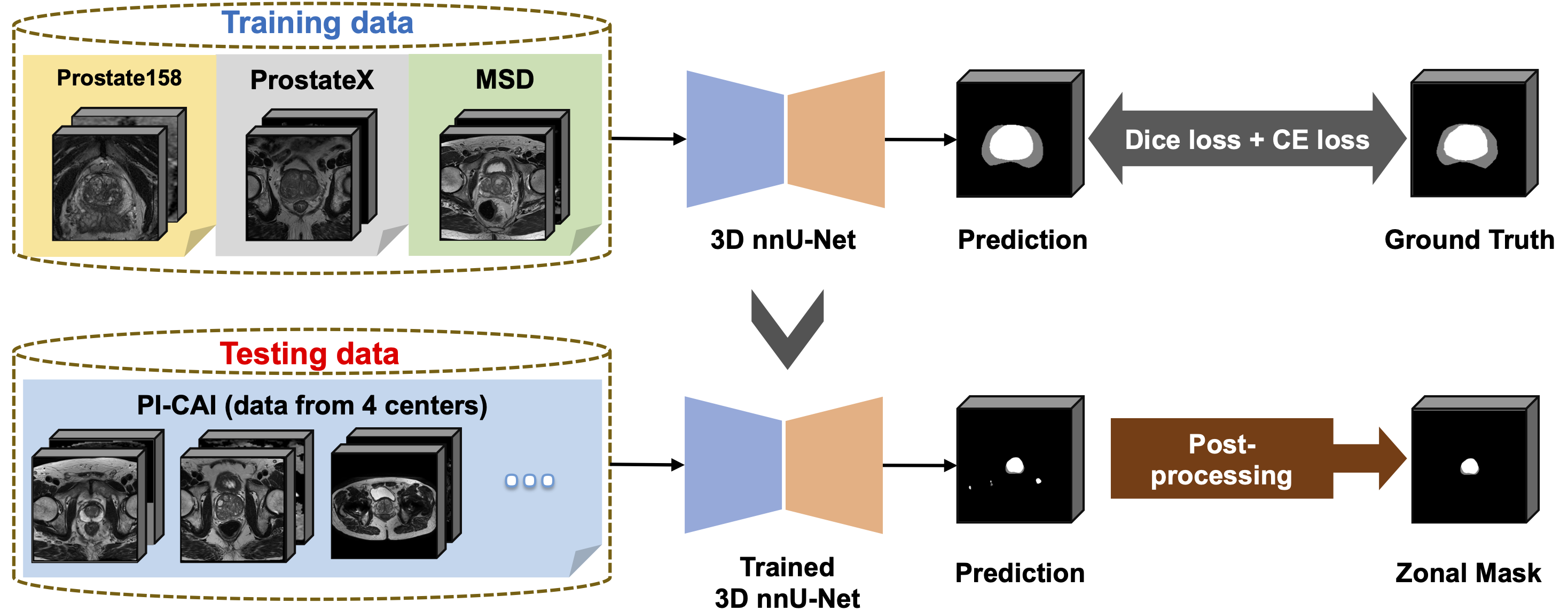}}
\caption{Overview of the zonal mask generation process. The T2W and ADC images are used for model training and testing.}
\label{figa1}
\end{figure*}

\begin{figure}[]
\centerline{\includegraphics[width=0.5\textwidth]{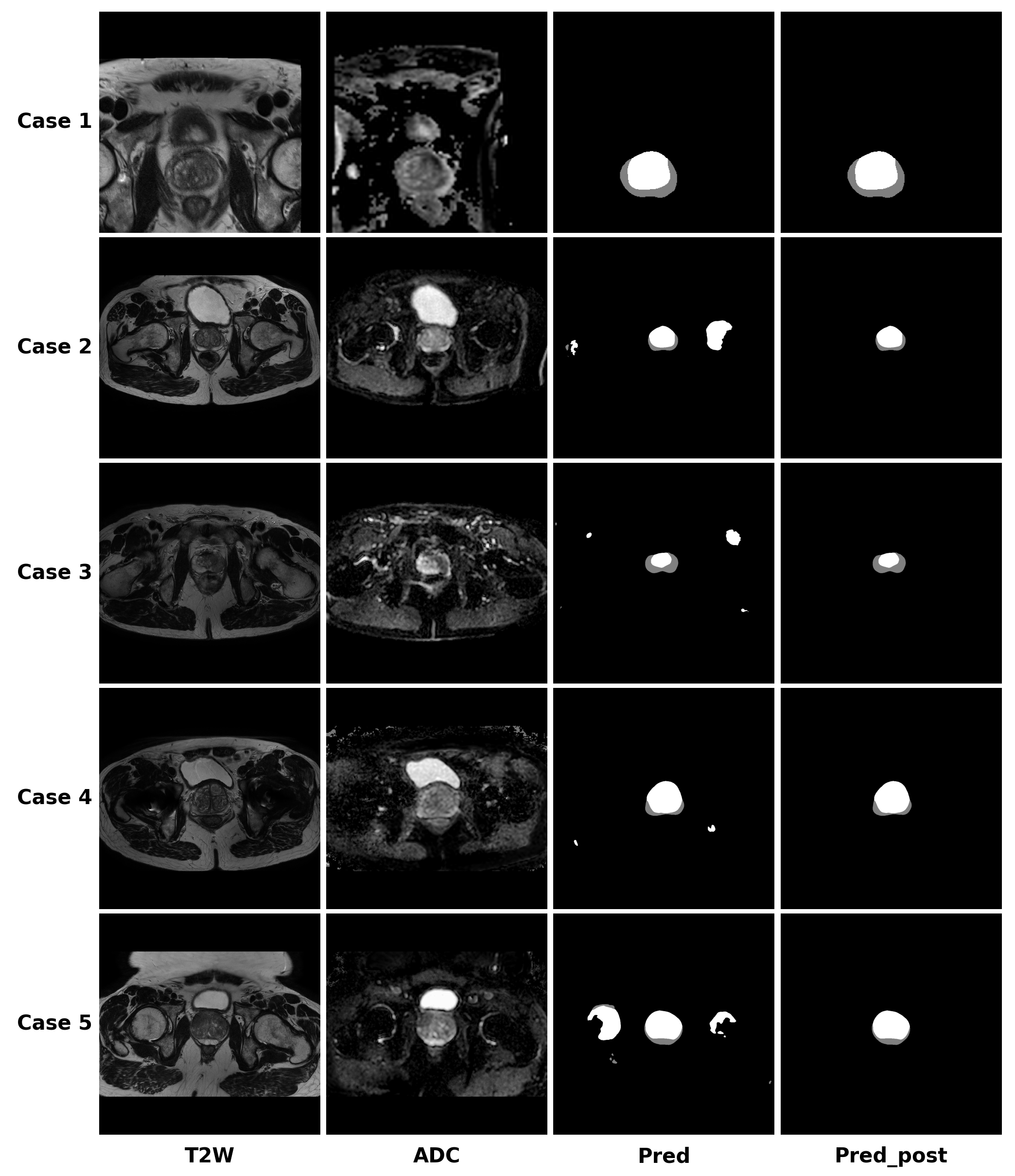}}
\caption{The examples of the generated zonal masks with and without post-processing.}
\label{figa2}
\end{figure}

%% Labels are used to cross-reference an item using \ref command.
We trained a 3D nnU-Net \cite{isensee2021nnu} model using T2W and ADC images with manual labels from 3 public datasets (Prostate158 dataset \cite{keno_bressem_2022_6481141}, ProstateX dataset \cite{litjens2017prostatex} and MSD dataset \cite{antonelli2022medical}) with 394 cases to generate prostate zonal segmentation masks (CG and PZ). The overview of this process is shown in Figure \ref{figa1}. The field of view (FOV) difference between the training datasets and the PI-CAI dataset results in poor model generalization. To address this, we removed noisy markers outside the prostate region using the following rules: 1) if there is only one connected component, keep it; 2) remove the connected components with only one kind of anatomical region; 3) keep the connected component in the middle closest to the midpoint of images with a large FOV and, 4) remove areas other than the largest connected component. Finally, we used the trained model to generate the zonal masks of the PI-CAI dataset. We provide several examples of the generated zonal masks with and without post-processing in Figure \ref{figa2}.

\section{Image transformations for SSL pre-training.}
\begin{figure}[]
\centerline{\includegraphics[width=0.5\textwidth]{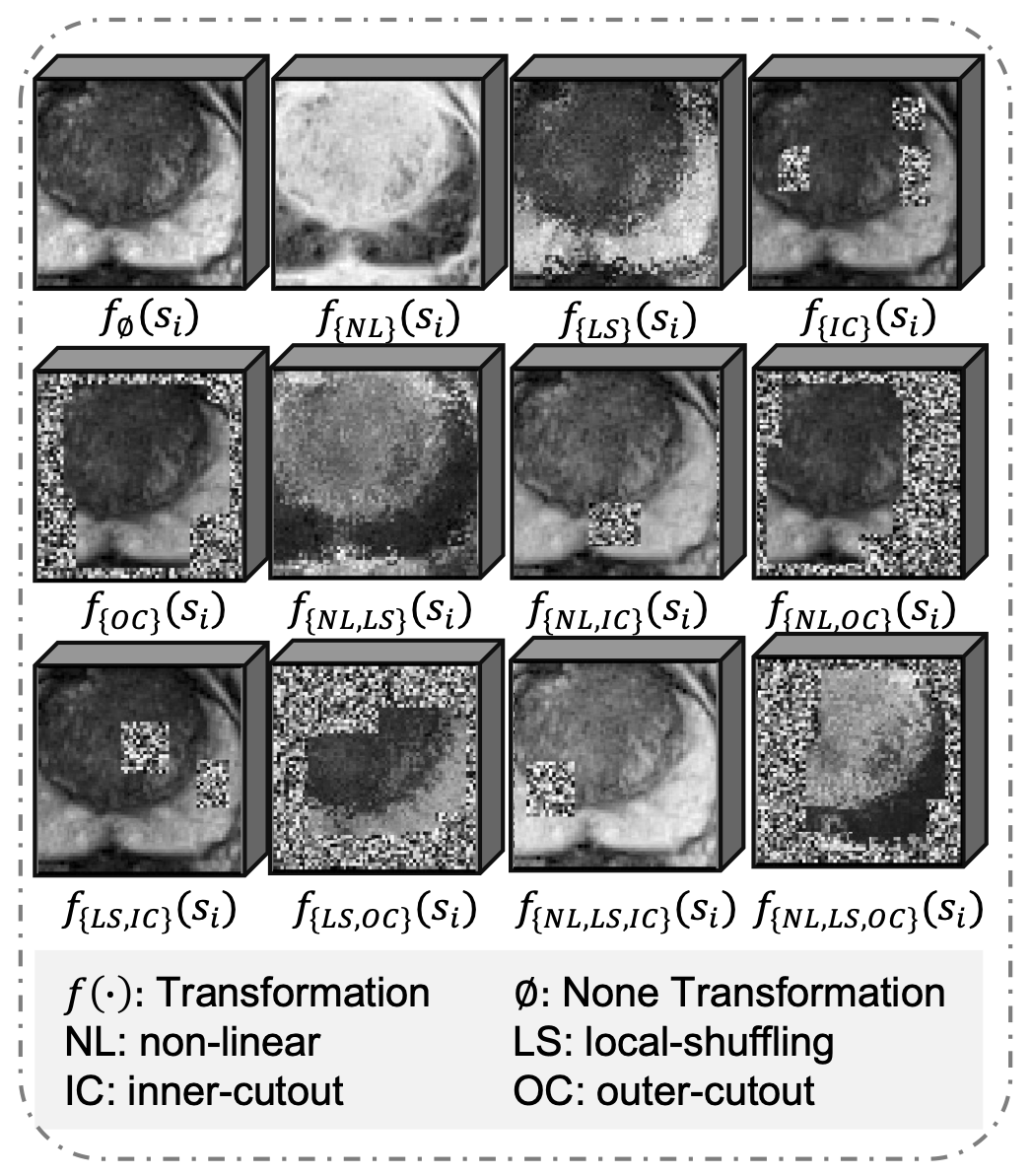}}
\caption{The image transformations used to corrupt sub-volumes. Each sub-volume $s_{i}$ can undergo at most three out of four transformations: non-linear, local-shuffling, inner-cutout, and outer-cutout, resulting in a transformed sub-volume $\widehat{s_{i}}$. It should be noted that inner-cutout and outer-cutout are considered mutually exclusive. Therefore, in addition to the four original individual transformations, this process yields eight more transformations, including one identity mapping $f(\phi)$(meaning none of the four individual transformations is selected) and seven combined transformations.}
\label{figb3}
\end{figure}

The illustration of image transformations used to corrupt extracted sub-volumes is shown in Figure \ref{figb3}.

\section{Cross-validation Performance Analyses}
\begin{figure*}[]
    \centering
    \begin{subfigure}[b]{0.32\textwidth}
        \includegraphics[width=\textwidth]{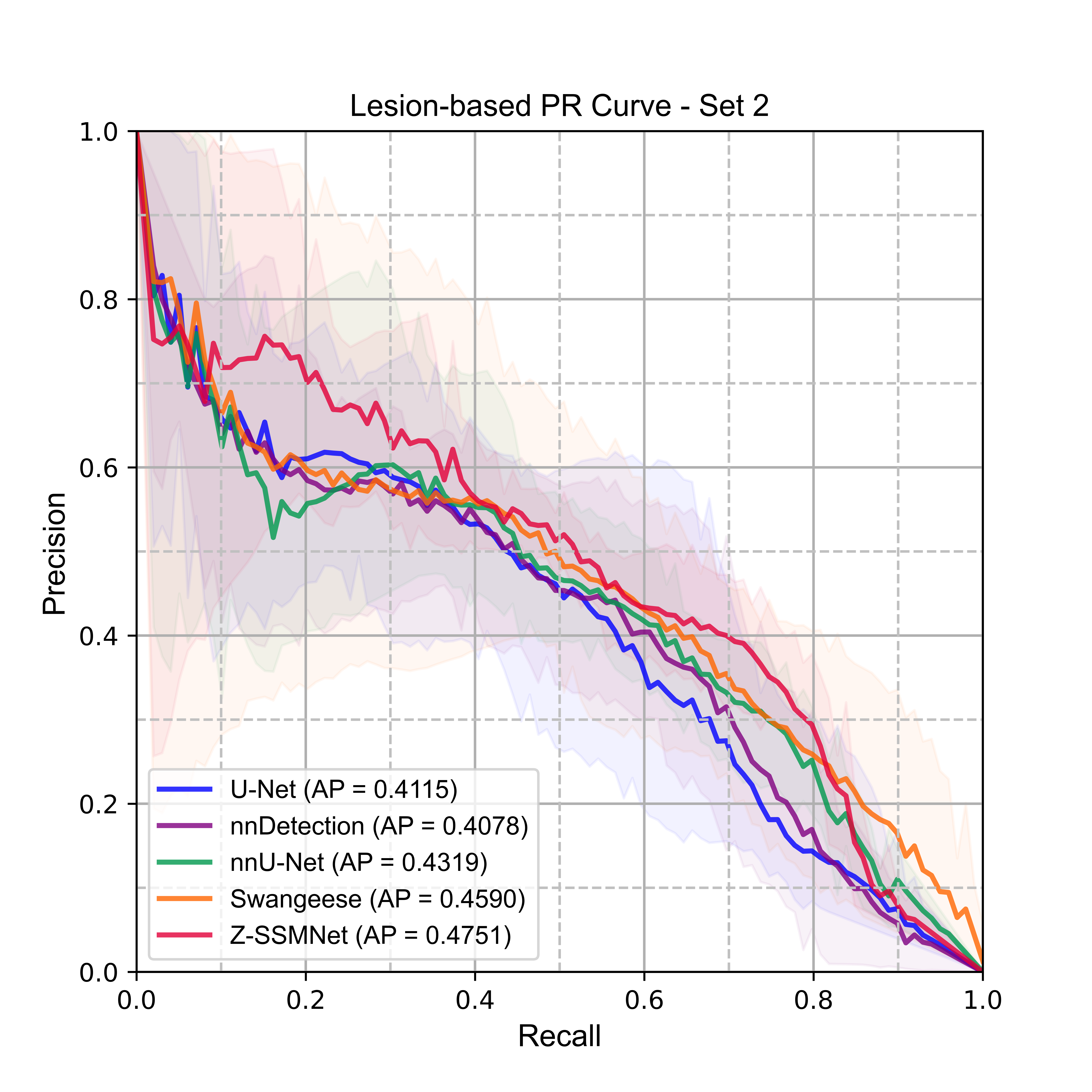}
        % \caption{}
        \label{fig:ap_weakly_supervised}
    \end{subfigure}
    % \hfill
    \begin{subfigure}[b]{0.32\textwidth}
        \includegraphics[width=\textwidth]{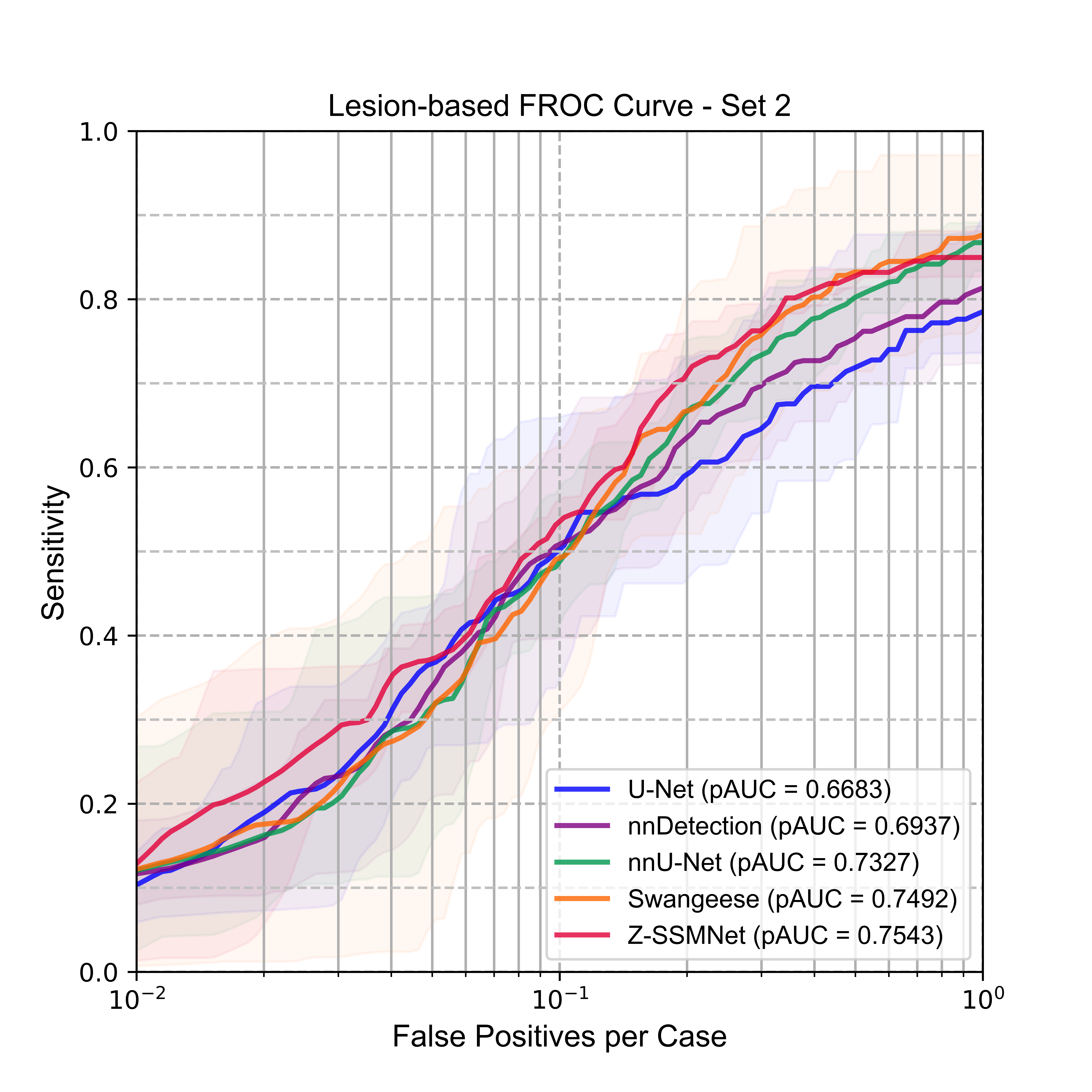}
        % \caption{}
    \label{fig:froc_weakly_supervised}
    \end{subfigure}
    \begin{subfigure}[b]{0.32\textwidth}
        \includegraphics[width=\textwidth]{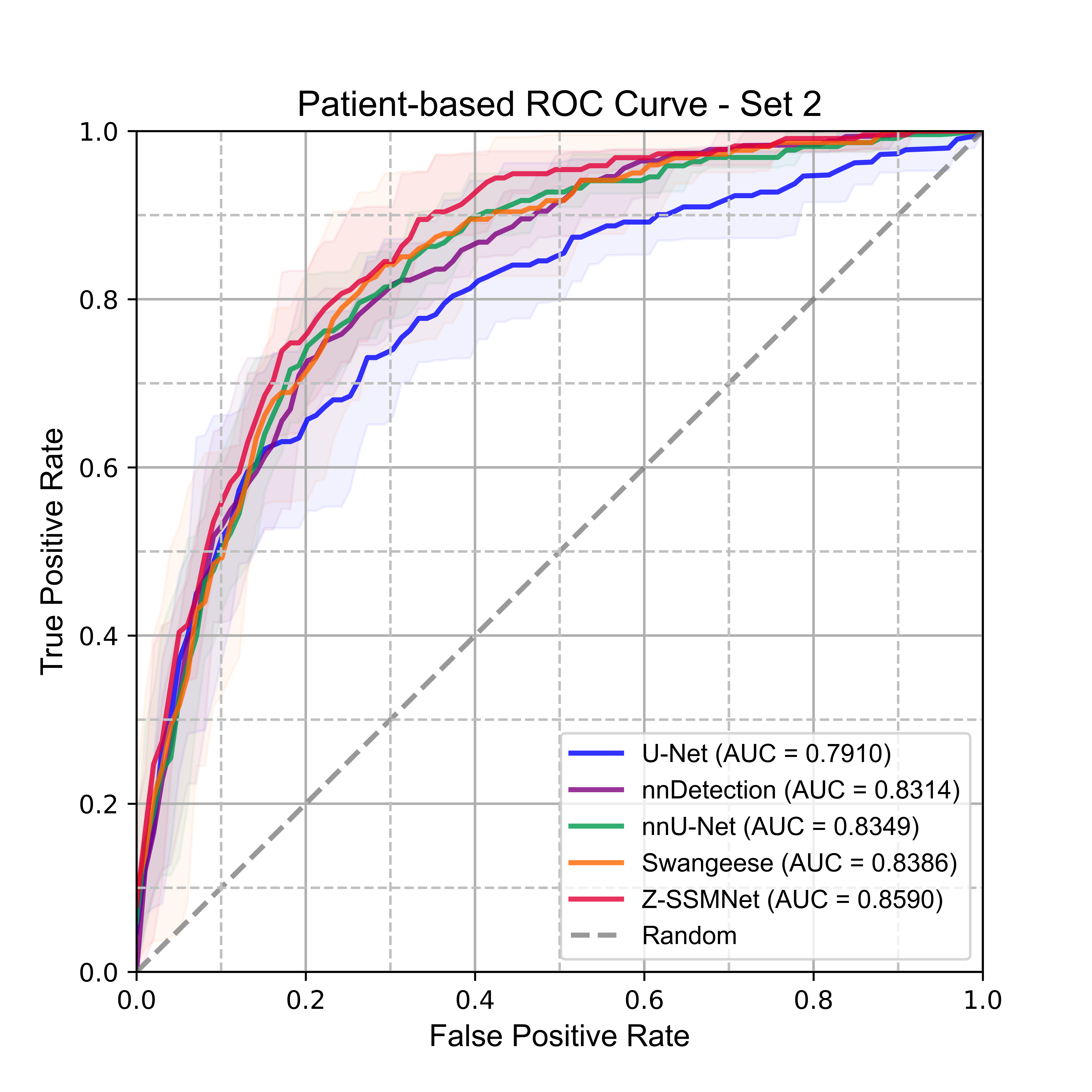}
        % \caption{}
        \label{fig:roc_weakly_supervised}
    \end{subfigure}
    % \hfill
\caption{Lesion-level PR, FROC and patient-level ROC analysis of methods trained on Set 2 of the PI-CAI dataset. Transparent areas indicate the 95\% confidence intervals.}
\label{figd4}
\end{figure*}

We visualize the lesion-level PR and FROC curves, along with the patient-level ROC curve of our Z-SSMNet, and compare it to baseline methods including 3D U-Net \cite{cciccek20163d}, nnU-Net \cite{isensee2021nnu} and nnDetection \cite{baumgartner2021nndetection}, and the method proposed by Kan et al. \cite{kanimplementation} (see Figure \ref{figd4}). All models are trained on Set 2 of the PI-CAI dataset. The comparison results demonstrate that our Z-SSMNet outperforms the other methods.

\bibliographystyle{elsarticle-num-names}
\bibliography{manuscript}

\begin{thebibliography}{45}
\expandafter\ifx\csname natexlab\endcsname\relax\def\natexlab#1{#1}\fi
\providecommand{\url}[1]{\texttt{#1}}
\providecommand{\href}[2]{#2}
\providecommand{\path}[1]{#1}
\providecommand{\DOIprefix}{doi:}
\providecommand{\ArXivprefix}{arXiv:}
\providecommand{\URLprefix}{URL: }
\providecommand{\Pubmedprefix}{pmid:}
\providecommand{\doi}[1]{\href{http://dx.doi.org/#1}{\path{#1}}}
\providecommand{\Pubmed}[1]{\href{pmid:#1}{\path{#1}}}
\providecommand{\bibinfo}[2]{#2}
\ifx\xfnm\relax \def\xfnm[#1]{\unskip,\space#1}\fi
%Type = Article
\bibitem[{Bray et~al.(2024)Bray, Laversanne, Sung, Ferlay, Siegel, Soerjomataram, and Jemal}]{bray2024global}
\bibinfo{author}{F.~Bray}, \bibinfo{author}{M.~Laversanne}, \bibinfo{author}{H.~Sung}, \bibinfo{author}{J.~Ferlay}, \bibinfo{author}{R.~L. Siegel}, \bibinfo{author}{I.~Soerjomataram}, \bibinfo{author}{A.~Jemal},
\newblock \bibinfo{title}{Global cancer statistics 2022: Globocan estimates of incidence and mortality worldwide for 36 cancers in 185 countries},
\newblock \bibinfo{journal}{CA: a cancer journal for clinicians} \bibinfo{volume}{74} (\bibinfo{year}{2024}) \bibinfo{pages}{229--263}.
%Type = Article
\bibitem[{Moses et~al.(2023)Moses, Sprenkle, Bahler, Box, Carlsson, Catalona, Dahl, Dall’Era, Davis, Drake et~al.}]{moses2023prostate}
\bibinfo{author}{K.~A. Moses}, \bibinfo{author}{P.~C. Sprenkle}, \bibinfo{author}{C.~Bahler}, \bibinfo{author}{G.~Box}, \bibinfo{author}{S.~V. Carlsson}, \bibinfo{author}{W.~J. Catalona}, \bibinfo{author}{D.~M. Dahl}, \bibinfo{author}{M.~Dall’Era}, \bibinfo{author}{J.~W. Davis}, \bibinfo{author}{B.~F. Drake}, et~al.,
\newblock \bibinfo{title}{Prostate cancer early detection, version 1.2023 featured updates to the nccn guidelines},
\newblock \bibinfo{journal}{JNCCN Journal of the National Comprehensive Cancer Network} \bibinfo{volume}{21} (\bibinfo{year}{2023}) \bibinfo{pages}{236--246}.
%Type = Article
\bibitem[{Tamada et~al.(2021)Tamada, Kido, Yamamoto, Takeuchi, Miyaji, Moriya, and Sone}]{tamada2021comparison}
\bibinfo{author}{T.~Tamada}, \bibinfo{author}{A.~Kido}, \bibinfo{author}{A.~Yamamoto}, \bibinfo{author}{M.~Takeuchi}, \bibinfo{author}{Y.~Miyaji}, \bibinfo{author}{T.~Moriya}, \bibinfo{author}{T.~Sone},
\newblock \bibinfo{title}{Comparison of biparametric and multiparametric mri for clinically significant prostate cancer detection with pi-rads version 2.1},
\newblock \bibinfo{journal}{Journal of Magnetic Resonance Imaging} \bibinfo{volume}{53} (\bibinfo{year}{2021}) \bibinfo{pages}{283--291}.
%Type = Article
\bibitem[{Eklund et~al.(2021)Eklund, J{\"a}derling, Discacciati, Bergman, Annerstedt, Aly, Glaessgen, Carlsson, Gr{\"o}nberg, and Nordstr{\"o}m}]{eklund2021mri}
\bibinfo{author}{M.~Eklund}, \bibinfo{author}{F.~J{\"a}derling}, \bibinfo{author}{A.~Discacciati}, \bibinfo{author}{M.~Bergman}, \bibinfo{author}{M.~Annerstedt}, \bibinfo{author}{M.~Aly}, \bibinfo{author}{A.~Glaessgen}, \bibinfo{author}{S.~Carlsson}, \bibinfo{author}{H.~Gr{\"o}nberg}, \bibinfo{author}{T.~Nordstr{\"o}m},
\newblock \bibinfo{title}{Mri-targeted or standard biopsy in prostate cancer screening},
\newblock \bibinfo{journal}{New England journal of medicine} \bibinfo{volume}{385} (\bibinfo{year}{2021}) \bibinfo{pages}{908--920}.
%Type = Article
\bibitem[{Sharma et~al.(2023)Sharma, Nayak, Balabantaray, Tanveer, and Nayak}]{sharma2023survey}
\bibinfo{author}{P.~Sharma}, \bibinfo{author}{D.~R. Nayak}, \bibinfo{author}{B.~K. Balabantaray}, \bibinfo{author}{M.~Tanveer}, \bibinfo{author}{R.~Nayak},
\newblock \bibinfo{title}{A survey on cancer detection via convolutional neural networks: Current challenges and future directions},
\newblock \bibinfo{journal}{Neural Networks}  (\bibinfo{year}{2023}).
%Type = Article
\bibitem[{Iglesias et~al.(2024)Iglesias, Talavera, Troya, D{\'\i}az-{\'A}lvarez, and Garc{\'\i}a-Remesal}]{iglesias2024artificial}
\bibinfo{author}{G.~Iglesias}, \bibinfo{author}{E.~Talavera}, \bibinfo{author}{J.~Troya}, \bibinfo{author}{A.~D{\'\i}az-{\'A}lvarez}, \bibinfo{author}{M.~Garc{\'\i}a-Remesal},
\newblock \bibinfo{title}{Artificial intelligence model for tumoral clinical decision support systems},
\newblock \bibinfo{journal}{Computer Methods and Programs in Biomedicine} \bibinfo{volume}{253} (\bibinfo{year}{2024}) \bibinfo{pages}{108228}.
%Type = Article
\bibitem[{Sunoqrot et~al.(2022)Sunoqrot, Saha, Hosseinzadeh, Elschot, and Huisman}]{sunoqrot2022artificial}
\bibinfo{author}{M.~R. Sunoqrot}, \bibinfo{author}{A.~Saha}, \bibinfo{author}{M.~Hosseinzadeh}, \bibinfo{author}{M.~Elschot}, \bibinfo{author}{H.~Huisman},
\newblock \bibinfo{title}{Artificial intelligence for prostate mri: open datasets, available applications, and grand challenges},
\newblock \bibinfo{journal}{European Radiology Experimental} \bibinfo{volume}{6} (\bibinfo{year}{2022}) \bibinfo{pages}{35}.
%Type = Article
\bibitem[{Saha et~al.(2024)Saha, Bosma, Twilt, van Ginneken, Bjartell, Padhani, Bonekamp, Villeirs, Salomon, Giannarini et~al.}]{saha2024artificial}
\bibinfo{author}{A.~Saha}, \bibinfo{author}{J.~S. Bosma}, \bibinfo{author}{J.~J. Twilt}, \bibinfo{author}{B.~van Ginneken}, \bibinfo{author}{A.~Bjartell}, \bibinfo{author}{A.~R. Padhani}, \bibinfo{author}{D.~Bonekamp}, \bibinfo{author}{G.~Villeirs}, \bibinfo{author}{G.~Salomon}, \bibinfo{author}{G.~Giannarini}, et~al.,
\newblock \bibinfo{title}{Artificial intelligence and radiologists in prostate cancer detection on mri (pi-cai): an international, paired, non-inferiority, confirmatory study},
\newblock \bibinfo{journal}{The Lancet Oncology}  (\bibinfo{year}{2024}).
%Type = Article
\bibitem[{Wang et~al.(2018)Wang, Liu, Cheng, Wang, Yang, and Cheng}]{wang2018automated}
\bibinfo{author}{Z.~Wang}, \bibinfo{author}{C.~Liu}, \bibinfo{author}{D.~Cheng}, \bibinfo{author}{L.~Wang}, \bibinfo{author}{X.~Yang}, \bibinfo{author}{K.-T. Cheng},
\newblock \bibinfo{title}{Automated detection of clinically significant prostate cancer in mp-mri images based on an end-to-end deep neural network},
\newblock \bibinfo{journal}{IEEE transactions on medical imaging} \bibinfo{volume}{37} (\bibinfo{year}{2018}) \bibinfo{pages}{1127--1139}.
%Type = Article
\bibitem[{Cao et~al.(2019)Cao, Bajgiran, Mirak, Shakeri, Zhong, Enzmann, Raman, and Sung}]{cao2019joint}
\bibinfo{author}{R.~Cao}, \bibinfo{author}{A.~M. Bajgiran}, \bibinfo{author}{S.~A. Mirak}, \bibinfo{author}{S.~Shakeri}, \bibinfo{author}{X.~Zhong}, \bibinfo{author}{D.~Enzmann}, \bibinfo{author}{S.~Raman}, \bibinfo{author}{K.~Sung},
\newblock \bibinfo{title}{Joint prostate cancer detection and gleason score prediction in mp-mri via focalnet},
\newblock \bibinfo{journal}{IEEE transactions on medical imaging} \bibinfo{volume}{38} (\bibinfo{year}{2019}) \bibinfo{pages}{2496--2506}.
%Type = Article
\bibitem[{Duran et~al.(2022)Duran, Dussert, Rouvi{\`e}re, Jaouen, Jodoin, and Lartizien}]{duran2022prostattention}
\bibinfo{author}{A.~Duran}, \bibinfo{author}{G.~Dussert}, \bibinfo{author}{O.~Rouvi{\`e}re}, \bibinfo{author}{T.~Jaouen}, \bibinfo{author}{P.-M. Jodoin}, \bibinfo{author}{C.~Lartizien},
\newblock \bibinfo{title}{Prostattention-net: A deep attention model for prostate cancer segmentation by aggressiveness in mri scans},
\newblock \bibinfo{journal}{Medical Image Analysis} \bibinfo{volume}{77} (\bibinfo{year}{2022}) \bibinfo{pages}{102347}.
%Type = Article
\bibitem[{Turkbey et~al.(2019)Turkbey, Rosenkrantz, Haider, Padhani, Villeirs, Macura, Tempany, Choyke, Cornud, Margolis et~al.}]{turkbey2019prostate}
\bibinfo{author}{B.~Turkbey}, \bibinfo{author}{A.~B. Rosenkrantz}, \bibinfo{author}{M.~A. Haider}, \bibinfo{author}{A.~R. Padhani}, \bibinfo{author}{G.~Villeirs}, \bibinfo{author}{K.~J. Macura}, \bibinfo{author}{C.~M. Tempany}, \bibinfo{author}{P.~L. Choyke}, \bibinfo{author}{F.~Cornud}, \bibinfo{author}{D.~J. Margolis}, et~al.,
\newblock \bibinfo{title}{Prostate imaging reporting and data system version 2.1: 2019 update of prostate imaging reporting and data system version 2},
\newblock \bibinfo{journal}{European urology} \bibinfo{volume}{76} (\bibinfo{year}{2019}) \bibinfo{pages}{340--351}.
%Type = Article
\bibitem[{De~Vente et~al.(2020)De~Vente, Vos, Hosseinzadeh, Pluim, and Veta}]{de2020deep}
\bibinfo{author}{C.~De~Vente}, \bibinfo{author}{P.~Vos}, \bibinfo{author}{M.~Hosseinzadeh}, \bibinfo{author}{J.~Pluim}, \bibinfo{author}{M.~Veta},
\newblock \bibinfo{title}{Deep learning regression for prostate cancer detection and grading in bi-parametric mri},
\newblock \bibinfo{journal}{IEEE Transactions on Biomedical Engineering} \bibinfo{volume}{68} (\bibinfo{year}{2020}) \bibinfo{pages}{374--383}.
%Type = Article
\bibitem[{Zhu et~al.(2022)Zhu, Gao, Zhu, Han, Liu, Li, Liu, Wang, Zhang, Zhang et~al.}]{zhu2022fully}
\bibinfo{author}{L.~Zhu}, \bibinfo{author}{G.~Gao}, \bibinfo{author}{Y.~Zhu}, \bibinfo{author}{C.~Han}, \bibinfo{author}{X.~Liu}, \bibinfo{author}{D.~Li}, \bibinfo{author}{W.~Liu}, \bibinfo{author}{X.~Wang}, \bibinfo{author}{J.~Zhang}, \bibinfo{author}{X.~Zhang}, et~al.,
\newblock \bibinfo{title}{Fully automated detection and localization of clinically significant prostate cancer on mr images using a cascaded convolutional neural network},
\newblock \bibinfo{journal}{Frontiers in Oncology} \bibinfo{volume}{12} (\bibinfo{year}{2022}) \bibinfo{pages}{958065}.
%Type = Article
\bibitem[{Seetharaman et~al.(2021)Seetharaman, Bhattacharya, Chen, Kunder, Shao, Soerensen, Wang, Teslovich, Fan, Ghanouni et~al.}]{seetharaman2021automated}
\bibinfo{author}{A.~Seetharaman}, \bibinfo{author}{I.~Bhattacharya}, \bibinfo{author}{L.~C. Chen}, \bibinfo{author}{C.~A. Kunder}, \bibinfo{author}{W.~Shao}, \bibinfo{author}{S.~J. Soerensen}, \bibinfo{author}{J.~B. Wang}, \bibinfo{author}{N.~C. Teslovich}, \bibinfo{author}{R.~E. Fan}, \bibinfo{author}{P.~Ghanouni}, et~al.,
\newblock \bibinfo{title}{Automated detection of aggressive and indolent prostate cancer on magnetic resonance imaging},
\newblock \bibinfo{journal}{Medical Physics} \bibinfo{volume}{48} (\bibinfo{year}{2021}) \bibinfo{pages}{2960--2972}.
%Type = Article
\bibitem[{Adams et~al.(2022)Adams, Makowski, Engel, Rattunde, Busch, Asbach, Niehues, Vinayahalingam, van Ginneken, Litjens et~al.}]{adams2022prostate158}
\bibinfo{author}{L.~C. Adams}, \bibinfo{author}{M.~R. Makowski}, \bibinfo{author}{G.~Engel}, \bibinfo{author}{M.~Rattunde}, \bibinfo{author}{F.~Busch}, \bibinfo{author}{P.~Asbach}, \bibinfo{author}{S.~M. Niehues}, \bibinfo{author}{S.~Vinayahalingam}, \bibinfo{author}{B.~van Ginneken}, \bibinfo{author}{G.~Litjens}, et~al.,
\newblock \bibinfo{title}{Prostate158-an expert-annotated 3t mri dataset and algorithm for prostate cancer detection},
\newblock \bibinfo{journal}{Computers in Biology and Medicine} \bibinfo{volume}{148} (\bibinfo{year}{2022}) \bibinfo{pages}{105817}.
%Type = Article
\bibitem[{Bosma et~al.(2023)Bosma, Saha, Hosseinzadeh, Slootweg, de~Rooij, and Huisman}]{bosma2023semi}
\bibinfo{author}{J.~S. Bosma}, \bibinfo{author}{A.~Saha}, \bibinfo{author}{M.~Hosseinzadeh}, \bibinfo{author}{I.~Slootweg}, \bibinfo{author}{M.~de~Rooij}, \bibinfo{author}{H.~Huisman},
\newblock \bibinfo{title}{Semi-supervised learning with report-guided pseudo labels for deep learning-based prostate cancer detection using biparametric mri},
\newblock \bibinfo{journal}{Radiology: Artificial Intelligence}  (\bibinfo{year}{2023}) \bibinfo{pages}{e230031}.
%Type = Article
\bibitem[{Yu and Dai(2024)}]{yu2024self}
\bibinfo{author}{H.~Yu}, \bibinfo{author}{Q.~Dai},
\newblock \bibinfo{title}{Self-supervised multi-task learning for medical image analysis},
\newblock \bibinfo{journal}{Pattern Recognition}  (\bibinfo{year}{2024}) \bibinfo{pages}{110327}.
%Type = Article
\bibitem[{Frueh et~al.(2022)Frueh, Kuestner, Nachbar, Thorwarth, Schilling, and Gatidis}]{frueh2022self}
\bibinfo{author}{M.~Frueh}, \bibinfo{author}{T.~Kuestner}, \bibinfo{author}{M.~Nachbar}, \bibinfo{author}{D.~Thorwarth}, \bibinfo{author}{A.~Schilling}, \bibinfo{author}{S.~Gatidis},
\newblock \bibinfo{title}{Self-supervised learning for automated anatomical tracking in medical image data with minimal human labeling effort},
\newblock \bibinfo{journal}{Computer Methods and Programs in Biomedicine} \bibinfo{volume}{225} (\bibinfo{year}{2022}) \bibinfo{pages}{107085}.
%Type = Article
\bibitem[{Li et~al.(2021)Li, Zhao, Tao, Zhai, Chen, He, and Cai}]{li2021multi}
\bibinfo{author}{J.~Li}, \bibinfo{author}{G.~Zhao}, \bibinfo{author}{Y.~Tao}, \bibinfo{author}{P.~Zhai}, \bibinfo{author}{H.~Chen}, \bibinfo{author}{H.~He}, \bibinfo{author}{T.~Cai},
\newblock \bibinfo{title}{Multi-task contrastive learning for automatic ct and x-ray diagnosis of covid-19},
\newblock \bibinfo{journal}{Pattern recognition} \bibinfo{volume}{114} (\bibinfo{year}{2021}) \bibinfo{pages}{107848}.
%Type = Inproceedings
\bibitem[{Bolous et~al.(2021)Bolous, Seetharaman, Bhattacharya, Fan, Soerensen, Chen, Ghanouni, Sonn, and Rusu}]{bolous2021clinically}
\bibinfo{author}{A.~Bolous}, \bibinfo{author}{A.~Seetharaman}, \bibinfo{author}{I.~Bhattacharya}, \bibinfo{author}{R.~E. Fan}, \bibinfo{author}{S.~J.~C. Soerensen}, \bibinfo{author}{L.~Chen}, \bibinfo{author}{P.~Ghanouni}, \bibinfo{author}{G.~A. Sonn}, \bibinfo{author}{M.~Rusu},
\newblock \bibinfo{title}{Clinically significant prostate cancer detection on mri with self-supervised learning using image context restoration},
\newblock in: \bibinfo{booktitle}{Medical Imaging 2021: Computer-Aided Diagnosis}, volume \bibinfo{volume}{11597}, \bibinfo{organization}{SPIE}, \bibinfo{year}{2021}, pp. \bibinfo{pages}{382--387}.
%Type = Article
\bibitem[{Qian et~al.(2021)Qian, Zhang, and Wang}]{qian2021procdet}
\bibinfo{author}{Y.~Qian}, \bibinfo{author}{Z.~Zhang}, \bibinfo{author}{B.~Wang},
\newblock \bibinfo{title}{Procdet: a new method for prostate cancer detection based on mr images},
\newblock \bibinfo{journal}{IEEE Access} \bibinfo{volume}{9} (\bibinfo{year}{2021}) \bibinfo{pages}{143495--143505}.
%Type = Misc
\bibitem[{Saha et~al.(2022)Saha, Twilt, Bosma, van Ginneken, Yakar, Elschot, Veltman, Fütterer, de~Rooij, and Huisman}]{saha_2022_6517398}
\bibinfo{author}{A.~Saha}, \bibinfo{author}{J.~J. Twilt}, \bibinfo{author}{J.~S. Bosma}, \bibinfo{author}{B.~van Ginneken}, \bibinfo{author}{D.~Yakar}, \bibinfo{author}{M.~Elschot}, \bibinfo{author}{J.~Veltman}, \bibinfo{author}{J.~Fütterer}, \bibinfo{author}{M.~de~Rooij}, \bibinfo{author}{H.~Huisman}, \bibinfo{title}{{The PI-CAI Challenge: Public Training and Development Dataset}}, \bibinfo{year}{2022}. \URLprefix \url{https://doi.org/10.5281/zenodo.6517398}. \DOIprefix\doi{10.5281/zenodo.6517398}.
%Type = Misc
\bibitem[{Bressem et~al.(2022)Bressem, Adams, and Engel}]{keno_bressem_2022_6481141}
\bibinfo{author}{K.~Bressem}, \bibinfo{author}{L.~Adams}, \bibinfo{author}{G.~Engel}, \bibinfo{title}{Prostate158 - training data}, \bibinfo{year}{2022}. \URLprefix \url{https://doi.org/10.5281/zenodo.6481141}. \DOIprefix\doi{10.5281/zenodo.6481141}.
%Type = Article
\bibitem[{Litjens et~al.(2017)Litjens, Debats, Barentsz, Karssemeijer, and Huisman}]{litjens2017prostatex}
\bibinfo{author}{G.~Litjens}, \bibinfo{author}{O.~Debats}, \bibinfo{author}{J.~Barentsz}, \bibinfo{author}{N.~Karssemeijer}, \bibinfo{author}{H.~Huisman},
\newblock \bibinfo{title}{Spie-aapm prostatex challenge data (version 2) [dataset]},
\newblock \bibinfo{journal}{The cancer imaging archive} \bibinfo{volume}{10} (\bibinfo{year}{2017}) \bibinfo{pages}{K9TCIA}. \DOIprefix\doi{https://doi.org/10.7937/K9TCIA.2017.MURS5CL}.
%Type = Article
\bibitem[{Antonelli et~al.(2022)Antonelli, Reinke, Bakas, Farahani, Kopp-Schneider, Landman, Litjens, Menze, Ronneberger, Summers et~al.}]{antonelli2022medical}
\bibinfo{author}{M.~Antonelli}, \bibinfo{author}{A.~Reinke}, \bibinfo{author}{S.~Bakas}, \bibinfo{author}{K.~Farahani}, \bibinfo{author}{A.~Kopp-Schneider}, \bibinfo{author}{B.~A. Landman}, \bibinfo{author}{G.~Litjens}, \bibinfo{author}{B.~Menze}, \bibinfo{author}{O.~Ronneberger}, \bibinfo{author}{R.~M. Summers}, et~al.,
\newblock \bibinfo{title}{The medical segmentation decathlon},
\newblock \bibinfo{journal}{Nature communications} \bibinfo{volume}{13} (\bibinfo{year}{2022}) \bibinfo{pages}{4128}.
%Type = Article
\bibitem[{Cuocolo et~al.(2021)Cuocolo, Stanzione, Castaldo, De~Lucia, and Imbriaco}]{cuocolo2021quality}
\bibinfo{author}{R.~Cuocolo}, \bibinfo{author}{A.~Stanzione}, \bibinfo{author}{A.~Castaldo}, \bibinfo{author}{D.~R. De~Lucia}, \bibinfo{author}{M.~Imbriaco},
\newblock \bibinfo{title}{Quality control and whole-gland, zonal and lesion annotations for the prostatex challenge public dataset},
\newblock \bibinfo{journal}{European Journal of Radiology} \bibinfo{volume}{138} (\bibinfo{year}{2021}) \bibinfo{pages}{109647}.
%Type = Inproceedings
\bibitem[{Dong et~al.(2022)Dong, He, Qi, Chen, Shu, Coatrieux, Yang, and Li}]{dong2022mnet}
\bibinfo{author}{Z.~Dong}, \bibinfo{author}{Y.~He}, \bibinfo{author}{X.~Qi}, \bibinfo{author}{Y.~Chen}, \bibinfo{author}{H.~Shu}, \bibinfo{author}{J.-L. Coatrieux}, \bibinfo{author}{G.~Yang}, \bibinfo{author}{S.~Li},
\newblock \bibinfo{title}{Mnet: rethinking 2d/3d networks for anisotropic medical image segmentation},
\newblock in: \bibinfo{booktitle}{Thirty-First International Joint Conference on Artificial Intelligence $\{$IJCAI-22$\}$}, \bibinfo{organization}{International Joint Conferences on Artificial Intelligence Organization}, \bibinfo{year}{2022}, pp. \bibinfo{pages}{870--876}.
%Type = Book
\bibitem[{Mortenson(1999)}]{mortenson1999mathematics}
\bibinfo{author}{M.~E. Mortenson}, \bibinfo{title}{Mathematics for computer graphics applications}, \bibinfo{publisher}{Industrial Press Inc.}, \bibinfo{year}{1999}.
%Type = Inproceedings
\bibitem[{Pathak et~al.(2016)Pathak, Krahenbuhl, Donahue, Darrell, and Efros}]{pathak2016context}
\bibinfo{author}{D.~Pathak}, \bibinfo{author}{P.~Krahenbuhl}, \bibinfo{author}{J.~Donahue}, \bibinfo{author}{T.~Darrell}, \bibinfo{author}{A.~A. Efros},
\newblock \bibinfo{title}{Context encoders: Feature learning by inpainting},
\newblock in: \bibinfo{booktitle}{Proceedings of the IEEE conference on computer vision and pattern recognition}, \bibinfo{year}{2016}, pp. \bibinfo{pages}{2536--2544}.
%Type = Article
\bibitem[{Zhou et~al.(2021)Zhou, Sodha, Pang, Gotway, and Liang}]{zhou2021models}
\bibinfo{author}{Z.~Zhou}, \bibinfo{author}{V.~Sodha}, \bibinfo{author}{J.~Pang}, \bibinfo{author}{M.~B. Gotway}, \bibinfo{author}{J.~Liang},
\newblock \bibinfo{title}{Models genesis},
\newblock \bibinfo{journal}{Medical image analysis} \bibinfo{volume}{67} (\bibinfo{year}{2021}) \bibinfo{pages}{101840}.
%Type = Inproceedings
\bibitem[{Lin et~al.(2017)Lin, Goyal, Girshick, He, and Doll{\'a}r}]{lin2017focal}
\bibinfo{author}{T.-Y. Lin}, \bibinfo{author}{P.~Goyal}, \bibinfo{author}{R.~Girshick}, \bibinfo{author}{K.~He}, \bibinfo{author}{P.~Doll{\'a}r},
\newblock \bibinfo{title}{Focal loss for dense object detection},
\newblock in: \bibinfo{booktitle}{Proceedings of the IEEE international conference on computer vision}, \bibinfo{year}{2017}, pp. \bibinfo{pages}{2980--2988}.
%Type = Inproceedings
\bibitem[{{\c{C}}i{\c{c}}ek et~al.(2016){\c{C}}i{\c{c}}ek, Abdulkadir, Lienkamp, Brox, and Ronneberger}]{cciccek20163d}
\bibinfo{author}{{\"O}.~{\c{C}}i{\c{c}}ek}, \bibinfo{author}{A.~Abdulkadir}, \bibinfo{author}{S.~S. Lienkamp}, \bibinfo{author}{T.~Brox}, \bibinfo{author}{O.~Ronneberger},
\newblock \bibinfo{title}{3d u-net: learning dense volumetric segmentation from sparse annotation},
\newblock in: \bibinfo{booktitle}{Medical Image Computing and Computer-Assisted Intervention--MICCAI 2016: 19th International Conference, Athens, Greece, October 17-21, 2016, Proceedings, Part II 19}, \bibinfo{organization}{Springer}, \bibinfo{year}{2016}, pp. \bibinfo{pages}{424--432}.
%Type = Article
\bibitem[{Isensee et~al.(2021)Isensee, Jaeger, Kohl, Petersen, and Maier-Hein}]{isensee2021nnu}
\bibinfo{author}{F.~Isensee}, \bibinfo{author}{P.~F. Jaeger}, \bibinfo{author}{S.~A. Kohl}, \bibinfo{author}{J.~Petersen}, \bibinfo{author}{K.~H. Maier-Hein},
\newblock \bibinfo{title}{nnu-net: a self-configuring method for deep learning-based biomedical image segmentation},
\newblock \bibinfo{journal}{Nature methods} \bibinfo{volume}{18} (\bibinfo{year}{2021}) \bibinfo{pages}{203--211}.
%Type = Inproceedings
\bibitem[{Baumgartner et~al.(2021)Baumgartner, J{\"a}ger, Isensee, and Maier-Hein}]{baumgartner2021nndetection}
\bibinfo{author}{M.~Baumgartner}, \bibinfo{author}{P.~F. J{\"a}ger}, \bibinfo{author}{F.~Isensee}, \bibinfo{author}{K.~H. Maier-Hein},
\newblock \bibinfo{title}{nndetection: a self-configuring method for medical object detection},
\newblock in: \bibinfo{booktitle}{Medical Image Computing and Computer Assisted Intervention--MICCAI 2021: 24th International Conference, Strasbourg, France, September 27--October 1, 2021, Proceedings, Part V 24}, \bibinfo{organization}{Springer}, \bibinfo{year}{2021}, pp. \bibinfo{pages}{530--539}.
%Type = Article
\bibitem[{Debs et~al.(tion)Debs, Routier, Abi-Nader, Marcoux, Nicolas, B{\^o}ne, and Roh{\'e}}]{debsdeep}
\bibinfo{author}{N.~Debs}, \bibinfo{author}{A.~Routier}, \bibinfo{author}{C.~Abi-Nader}, \bibinfo{author}{A.~Marcoux}, \bibinfo{author}{F.~Nicolas}, \bibinfo{author}{A.~B{\^o}ne}, \bibinfo{author}{M.-M. Roh{\'e}},
\newblock \bibinfo{title}{Deep learning for detection and diagnosis of prostate cancer from bpmri and psa: Guerbet’s contribution to the pi-cai 2022 grand challenge}  (\bibinfo{year}{Personal communication}).
%Type = Article
\bibitem[{Kan et~al.(tion)Kan, Anhui, Qiao, Shi, and An}]{kanimplementation}
\bibinfo{author}{H.~Kan}, \bibinfo{author}{H.~Anhui}, \bibinfo{author}{L.~Qiao}, \bibinfo{author}{J.~Shi}, \bibinfo{author}{H.~An},
\newblock \bibinfo{title}{Implementation method of the pi-cai challenge (swangeese team)}  (\bibinfo{year}{Personal communication}).
%Type = Article
\bibitem[{Li et~al.(tion)Li, Vesal, Saunders, John, Soerensen, Jahanandish, Moroianu, Bhattacharya, Fan, Sonn et~al.}]{liprostate}
\bibinfo{author}{X.~Li}, \bibinfo{author}{S.~Vesal}, \bibinfo{author}{S.~Saunders}, \bibinfo{author}{S.~John}, \bibinfo{author}{C.~Soerensen}, \bibinfo{author}{H.~Jahanandish}, \bibinfo{author}{S.~Moroianu}, \bibinfo{author}{I.~Bhattacharya}, \bibinfo{author}{R.~E. Fan}, \bibinfo{author}{G.~A. Sonn}, et~al.,
\newblock \bibinfo{title}{The prostate imaging: Cancer ai (pi-cai) 2022 grand challenge (pimed team)}  (\bibinfo{year}{Personal communication}).
%Type = Article
\bibitem[{Karagoz et~al.(2023)Karagoz, Alis, Seker, Zeybel, Yergin, Oksuz, and Karaarslan}]{karagoz2023anatomically}
\bibinfo{author}{A.~Karagoz}, \bibinfo{author}{D.~Alis}, \bibinfo{author}{M.~E. Seker}, \bibinfo{author}{G.~Zeybel}, \bibinfo{author}{M.~Yergin}, \bibinfo{author}{I.~Oksuz}, \bibinfo{author}{E.~Karaarslan},
\newblock \bibinfo{title}{Anatomically guided self-adapting deep neural network for clinically significant prostate cancer detection on bi-parametric mri: a multi-center study},
\newblock \bibinfo{journal}{Insights into Imaging} \bibinfo{volume}{14} (\bibinfo{year}{2023}) \bibinfo{pages}{110}.
%Type = Inproceedings
\bibitem[{Jaeger et~al.(2020)Jaeger, Kohl, Bickelhaupt, Isensee, Kuder, Schlemmer, and Maier-Hein}]{jaeger2020retina}
\bibinfo{author}{P.~F. Jaeger}, \bibinfo{author}{S.~A. Kohl}, \bibinfo{author}{S.~Bickelhaupt}, \bibinfo{author}{F.~Isensee}, \bibinfo{author}{T.~A. Kuder}, \bibinfo{author}{H.-P. Schlemmer}, \bibinfo{author}{K.~H. Maier-Hein},
\newblock \bibinfo{title}{Retina u-net: Embarrassingly simple exploitation of segmentation supervision for medical object detection},
\newblock in: \bibinfo{booktitle}{Machine Learning for Health Workshop}, \bibinfo{organization}{PMLR}, \bibinfo{year}{2020}, pp. \bibinfo{pages}{171--183}.
%Type = Inproceedings
\bibitem[{Kan et~al.(2022)Kan, Shi, Zhao, Wang, Han, An, Wang, and Wang}]{kan2022itunet}
\bibinfo{author}{H.~Kan}, \bibinfo{author}{J.~Shi}, \bibinfo{author}{M.~Zhao}, \bibinfo{author}{Z.~Wang}, \bibinfo{author}{W.~Han}, \bibinfo{author}{H.~An}, \bibinfo{author}{Z.~Wang}, \bibinfo{author}{S.~Wang},
\newblock \bibinfo{title}{Itunet: Integration of transformers and unet for organs-at-risk segmentation},
\newblock in: \bibinfo{booktitle}{2022 44th Annual International Conference of the IEEE Engineering in Medicine \& Biology Society (EMBC)}, \bibinfo{organization}{IEEE}, \bibinfo{year}{2022}, pp. \bibinfo{pages}{2123--2127}.
%Type = Inproceedings
\bibitem[{Tan and Le(2019)}]{tan2019efficientnet}
\bibinfo{author}{M.~Tan}, \bibinfo{author}{Q.~Le},
\newblock \bibinfo{title}{Efficientnet: Rethinking model scaling for convolutional neural networks},
\newblock in: \bibinfo{booktitle}{International conference on machine learning}, \bibinfo{organization}{PMLR}, \bibinfo{year}{2019}, pp. \bibinfo{pages}{6105--6114}.
%Type = Article
\bibitem[{Saha et~al.(2021)Saha, Hosseinzadeh, and Huisman}]{saha2021end}
\bibinfo{author}{A.~Saha}, \bibinfo{author}{M.~Hosseinzadeh}, \bibinfo{author}{H.~Huisman},
\newblock \bibinfo{title}{End-to-end prostate cancer detection in bpmri via 3d cnns: effects of attention mechanisms, clinical priori and decoupled false positive reduction},
\newblock \bibinfo{journal}{Medical image analysis} \bibinfo{volume}{73} (\bibinfo{year}{2021}) \bibinfo{pages}{102155}.
%Type = Article
\bibitem[{Chen et~al.(2017)Chen, Papandreou, Kokkinos, Murphy, and Yuille}]{chen2017deeplab}
\bibinfo{author}{L.-C. Chen}, \bibinfo{author}{G.~Papandreou}, \bibinfo{author}{I.~Kokkinos}, \bibinfo{author}{K.~Murphy}, \bibinfo{author}{A.~L. Yuille},
\newblock \bibinfo{title}{Deeplab: Semantic image segmentation with deep convolutional nets, atrous convolution, and fully connected crfs},
\newblock \bibinfo{journal}{IEEE transactions on pattern analysis and machine intelligence} \bibinfo{volume}{40} (\bibinfo{year}{2017}) \bibinfo{pages}{834--848}.
%Type = Article
\bibitem[{Giannini et~al.(2015)Giannini, Vignati, De~Luca, Mazzetti, Russo, Armando, Stasi, Bollito, Porpiglia, and Regge}]{giannini2015novel}
\bibinfo{author}{V.~Giannini}, \bibinfo{author}{A.~Vignati}, \bibinfo{author}{M.~De~Luca}, \bibinfo{author}{S.~Mazzetti}, \bibinfo{author}{F.~Russo}, \bibinfo{author}{E.~Armando}, \bibinfo{author}{M.~Stasi}, \bibinfo{author}{E.~Bollito}, \bibinfo{author}{F.~Porpiglia}, \bibinfo{author}{D.~Regge},
\newblock \bibinfo{title}{A novel and fully automated registration method for prostate cancer detection using multiparametric magnetic resonance imaging},
\newblock \bibinfo{journal}{Journal of Medical Imaging and Health Informatics} \bibinfo{volume}{5} (\bibinfo{year}{2015}) \bibinfo{pages}{1171--1182}.

\end{thebibliography}
\end{document}